\documentclass[preprint,superscriptaddress,amsmath,amssymb,floatfix]{revtex4}

\usepackage{csquotes}
\usepackage{array}
\usepackage{graphicx}
\usepackage{extarrows}
\usepackage{url}
\usepackage{varioref}
\usepackage{hyperref}
\usepackage[super]{nth}
\usepackage{cleveref}

\usepackage{csquotes}
\usepackage{xcolor}
\usepackage[normalem]{ulem}
\usepackage{verbatim}

\begin{document}

%\title{Phase Doubling and Entanglement Generation in Coherent Many-Body Chemical Reactions}

\title{Observation of Phase Doubling and Entanglement in Coherent Matter-Wave Reactions}
\author{Shu Nagata}
\author{Tadej Me{\v z}nar{\v s}i{\v c}}
\author{Chuixin Kong}
\author{Cheng Chin$^*$}

\affiliation{James Franck Institute, Enrico Fermi Institute and Department of Physics, University of Chicago, Chicago, Illinois 60637, USA}

\maketitle 
%\linenumbers

{\bf Chemical reactions in a statistical ensemble are conventionally regarded as incoherent processes driven by thermodynamics. In the quantum degenerate regime, where atoms and molecules form coherent matter waves, reactions are theoretically described by nonlinear mixing of matter-wave fields. In this scenario, we expect phase matching between reactants and products, analogous to the mixing of photonic fields in nonlinear optics. Here we report on the observation of phase coherent reaction dynamics of Bose-condensed atoms and molecules near a Feshbach resonance. Using matter-wave diffraction with optical lattices, we verify spatial coherence of both atoms and molecules and observe phase doubling when atomic waves combine into molecular waves, the matter-wave analogue of optical frequency doubling. The diffraction patterns further reveal two-atom entanglement generated during the reaction. Our observations establish phase coherence and entanglement generation as two essential features of \enquote{quantum many-body chemistry}. Moreover, our work opens a pathway to control of reaction dynamics by manipulation of matter-wave phases.}

The wave-like behavior of massive particles, first predicted by Louis de Broglie ~\cite{Broglie1925}, established phase coherence as a central concept of quantum mechanics. In this picture, matter is described by a complex amplitude whose phase governs the dynamical evolution of the system and can be probed via  matter-wave interference.

This wave nature was first confirmed with electrons~\cite{Davisson1927}, and followed by atoms and molecules~\cite{Estermann1930}. Modern diffraction studies with large organic molecules~\cite{Brand2020} continue to test the limits of quantum coherence and its crossover to classical behavior. Control and measurement of matter-wave phases has led to powerful precision technologies, including neutron and atom interferometry~\cite{Greenberger1983, Bongs2019} and quantum sensing~\cite{Stray2022}.

Recent advances in cooling and trapping~\cite{Langen2024} now extend coherent control to molecules. Long molecular coherence times~\cite{Park2017, Zhou2020, Burchesky2021} and the generation of entanglement between molecular spins~\cite{Holland2023, Bao2023, Ruttley2025} establish ultracold molecules as a versatile platform for quantum information processing~\cite{Cornish2024}, matter-wave interferometry~\cite{Schmiedmayer} and controlled quantum chemistry~\cite{Bohn2017,Croft2017,Wolf2017,Chen2024,Liu2024,Martins2025}. These developments raise a natural question: how does matter-wave coherence evolve when particles undergo a chemical reaction?

The recent realization of Bose–Einstein condensates (BEC) of short‑range molecules~\cite{Zhang2021, Bigagli2024}, where a macroscopic ensemble of molecules shares a common phase, provides an ideal setting to answer this question. In the quantum degenerate regime, reactions are theoretically described by nonlinear mixing of matter-wave fields~\cite{Moore2002, Malla2022}, which features Bose-enhanced reactions and coherent matter-wave amplification - a phenomenon known as \enquote{quantum super-chemistry}~\cite{Heinzen2000}. A recent experiment with Cs$_2$ BECs~\cite{Zhang2023} supports this framework by demonstrating Bose-enhanced reactions and sets the stage to explore how coherence transforms during reactions.

\begin{figure}[t!]
    \centering
    \includegraphics[width=0.5\textwidth]{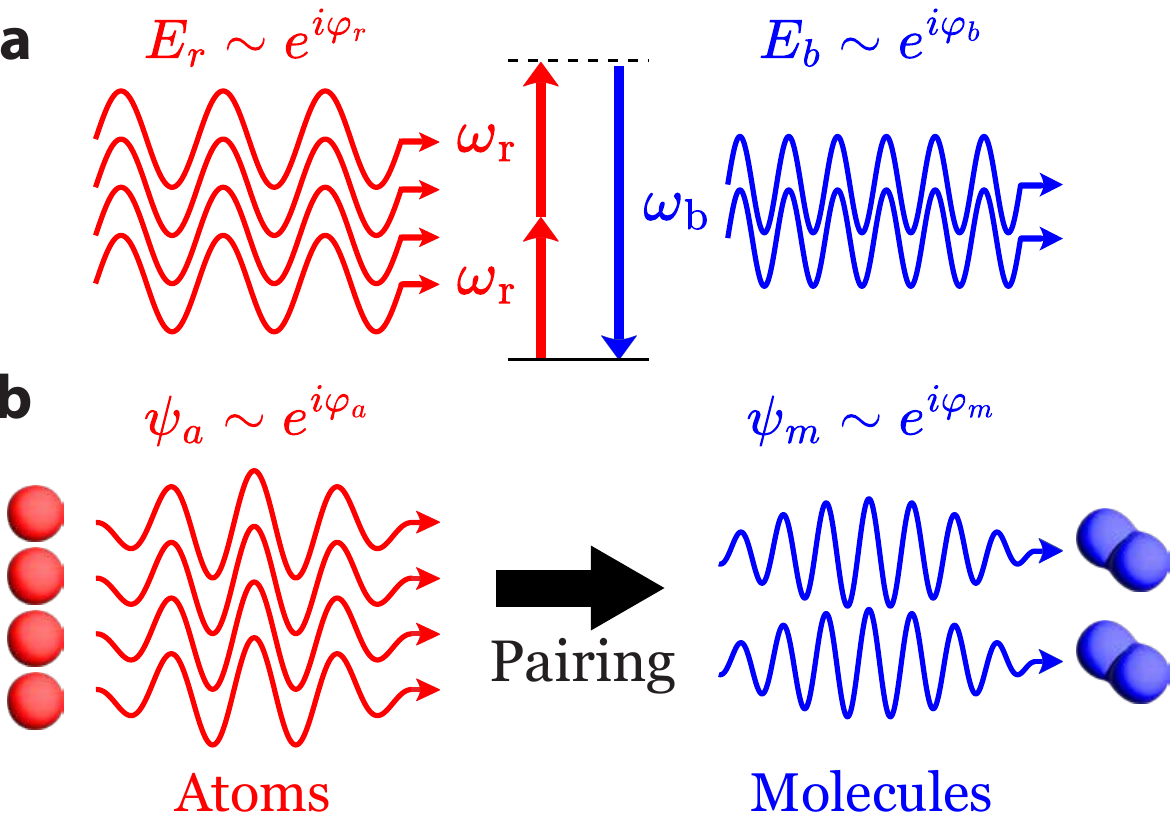}
    \caption{\textbf{Pairing atoms into molecules: matter-wave analogue of second harmonic generation in nonlinear optics.} \textbf{a}, Phase doubling in a nonlinear crystal $\varphi_\mathrm{b}=2\varphi_\mathrm{r}$. Pairs of red photons with momentum $k_\mathrm{r}=\nabla \varphi_\mathrm{r}$ and frequency $\omega_\mathrm{r}=-\partial_t \varphi_\mathrm{r}$ are converted into blue photons with twice the momentum $k_\mathrm{b}=2k_\mathrm{r}$ and frequency $
    \omega_\mathrm{b} = 2\omega_\mathrm{r}$. \textbf{b}, In the matter-wave analogue, atoms (red spheres) with wave function $\psi_\mathrm{a}$ are converted into diatomic molecules (blue spheres) with wave function $\psi_\mathrm{m}$. We show that phase doubling $\varphi_\mathrm{m}=2\varphi_\mathrm{a}$ and entanglement generation are key features in the nonlinear matter-wave mixing process.}
    \label{fig:Fig1}
\end{figure}

To illustrate how chemical reactions in a quantum gas can be understood as a nonlinear field mixing process, we consider the formation of molecules in a single quantum state $\psi_\mathrm{m}$ from pairs of atoms occupying the same state $\psi_\mathrm{a}$. This reaction is the matter-wave analogue of second-harmonic generation in nonlinear optics, where two red photons with frequency $\omega_\mathrm{r}$ are converted into a blue photon with frequency $\omega_\mathrm{b}=2\omega_\mathrm{r}$. The following interaction Hamiltonians describe the up- and down-conversion processes for photons, and the molecular association and dissociation processes for atomic fields,

\begin{align}
   &\mbox{Photon up/down conversion:}\; && \chi \hat{E}_\mathrm{b}^\dagger \hat{E}_\mathrm{r}^2 +  \chi \hat{E}_\mathrm{r}^{\dagger}{}^2 \hat{E}_\mathrm{b}, \\
   %&\mbox{Atoms:}\;\; i\hbar\partial_t\hat{\psi}_\mathrm{m}=\hbar\gamma\hat{\psi}_\mathrm{a}^2,
   &\mbox{Atom + atom $\rightleftarrows$ molecule:}\;  && \gamma\hat{\psi}_\mathrm{m}^\dagger \hat{\psi}_\mathrm{a}^2 + \gamma \hat{\psi}_\mathrm{a}^{\dagger}{}^2 \hat{\psi}_\mathrm{m}. \label{eq2:atom_hamiltonian}
\end{align}

\begin{figure*}[ht!]
    \centering
    \includegraphics[width=\textwidth]{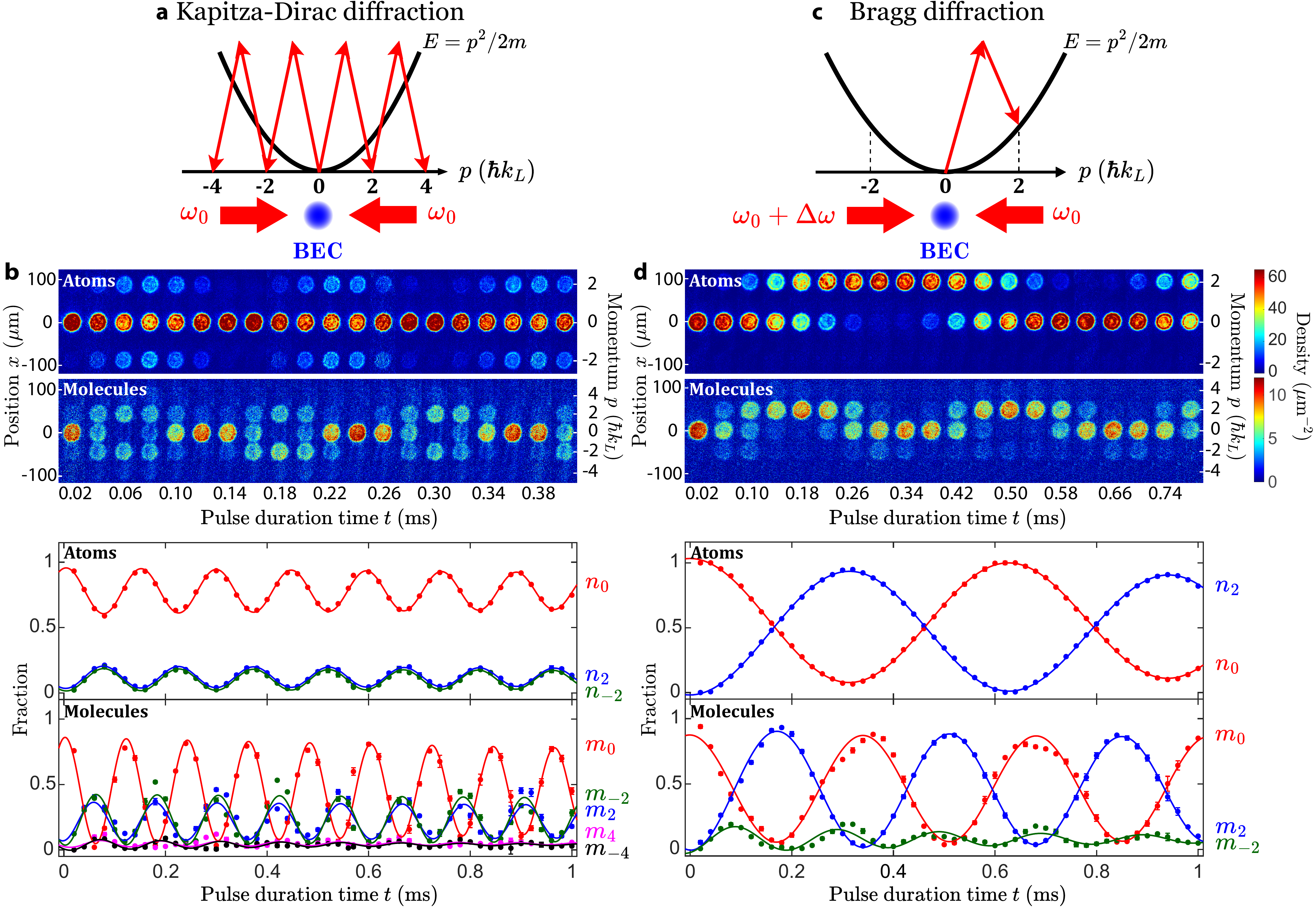}
    \caption{\textbf{Kapitza-Dirac and Bragg diffraction of atomic and molecular BECs.} 
    \textbf{a}, Kapitza-Dirac diffraction is induced by a pulse of counter-propagating laser beams with identical frequencies $\omega_{0}$ (wavelength is $\lambda=1064$~nm) for a duration of time $t$. The optical lattice potential diffracts particles at zero momentum $k=0$ to $\pm 2 k_L$, $\pm 4 k_L$..., where $k_L=2\pi/\lambda$ is the lattice momentum.  \textbf{b}, Rabi oscillations of the atoms and molecules in the momentum space for Kapitza-Dirac diffraction. Images are taken after a 15~ms time-of-flight. Here $n_j$ and $m_j$ denote the population fractions of atoms and molecules with momentum $j k_L$, respectively. The solid lines are sinusoidal fits which yield the Rabi frequencies $2\pi\times 6.77(1)$~kHz for atoms and $2\pi\times 8.31(1)$~kHz for molecules with momentum $k=0$ and $\pm2 k_L$. \textbf{c}, Bragg diffraction is realized by detuning one of the lattice beams by $\Delta\omega=4E_\mathrm{R}/\hbar=2\pi\times 5.304$~kHz for atoms and $2E_\mathrm{R}/\hbar=2\pi\times2.652$~kHz for molecules. The pulse resonantly scatters particles to momentum $k = 2 k_L$. Here, $E_\mathrm{R}=\hbar^2 k_L^2/2m$ is the atomic recoil energy and $m$ is the atomic mass. \textbf{d}, Rabi oscillations of the atoms and molecules for Bragg diffraction. The solid lines are sinusoidal fits which yield the Rabi frequency $2\pi\times1.59(1)$~kHz for atoms and $2\pi\times 2.95(1)$~kHz for molecules. Decoherence rates of the observed oscillations are lower than $0.2~\mathrm{ms}^{-1}$. Off-resonant excitations to momentum $k=-2k_L$ are stronger for molecules compared to atoms due to the larger Rabi coupling and smaller detuning~\cite{Supplement}. Images shown here are averaged over five independent measurements. Error bars show 1-$\sigma$ standard deviations of the means.}
    \label{fig:2}
\end{figure*}

\noindent where the coupling constant $\chi$ for photons is given by the second order susceptibility $\chi^{(2)}$~\cite{Agarwal1969}, and the coupling constant $\gamma$ for atoms can be induced by Feshbach resonances in cold collisions~\cite{Chin2010}. Here, $\hat{E}_\mathrm{r}$ and $\hat{E}_\mathrm{b}$ are the field operators of the red and blue photonic fields, and $\hat{\psi}_\mathrm{a}$ and $\hat{\psi}_\mathrm{m}$ are the field operators of the atomic and molecular fields.

An essential characteristic of nonlinear wave mixing in quantum optics is the phase matching of photonic fields. 
Similarly, the phase matching of matter-wave fields is predicted to occur when atoms are paired into molecules. To see this, we note that the Hamiltonian is symmetric with respect to the phase transformation $\hat{\psi}_\mathrm{a}\rightarrow\hat{\psi}_\mathrm{a}e^{i\varphi_\mathrm{a}}$ and $\hat{\psi}_\mathrm{m}\rightarrow\hat{\psi}_\mathrm{m}e^{i\varphi_m}=\hat{\psi}_\mathrm{m}e^{2i\varphi_\mathrm{a}}$, where $\varphi_\mathrm{a}$ and $\varphi_\mathrm{m}$ are the atomic and molecular phases, respectively~\cite{Zhang2023}. Thus phase doubling $\varphi_\mathrm{m}=2\varphi_\mathrm{a}$ preserves the reaction dynamics. Describing the atomic and molecular fields as coherent states, we can approximate the field operators with complex numbers, $\hat{\psi}_\mathrm{a}\rightarrow\sqrt{N}e^{i\varphi_\mathrm{a}}$ and $\hat{\psi}_\mathrm{m}\rightarrow\sqrt{M}e^{i\varphi_\mathrm{m}}$, where $N$ and $M$ are the atom and molecule number. The phase difference (reaction phase) $\varphi^{}_\mathrm{R}\equiv2\varphi_\mathrm{a}-\varphi_\mathrm{m}$ determines the direction of the chemical reaction, analogous to supercurrent in a Josephson junction~\cite{Supplement}

\begin{equation*}
   \dot{M}=-\frac{\dot{N}}2=2\gamma {N\sqrt{M}}  \sin\varphi^{}_\mathrm{R}.
\end{equation*}

\noindent When the atom and molecule populations reach an equilibrium $\dot{M}=-\dot{N}/2=0$, the phase is exactly doubled $\varphi_\mathrm{m}=2\varphi_\mathrm{a}$. 

Nonlinear field mixing is also a key process to generate entanglement between photons in quantum optics. %~\tm{\cite{gerry2023}}. 
Spontaneous parametric down conversion is a standard scheme to generate entangled photon pairs \cite{Anwar2021}. 
Continuous variable entanglement of twin beams is also routinely realized in optical parametric oscillators \cite{DellAnno2006, Adesso2007}.
%Pairing is analogous to the conjugate process, second harmonic generation.
Therefore, we also expect to see entanglement generation in the matter-wave analogue where atomic and molecular fields are reactively coupled \cite{Poulsen2001}. Reaction-induced entanglement may provide a powerful resource for realizing quantum control of chemical reactions~\cite{Gong2003,Li2019}.

In this paper, we report on phase doubling and entanglement generation when Bose-condensed atoms are paired into molecules. We first confirm that coherence is transferred from atoms to molecules during the reaction via matter‑wave diffraction. To demonstrate phase doubling, we imprint a phase pattern on the atomic BEC with an optical standing wave pulse. After pairing the atoms into molecules, we extract the phase of the molecular matter wave from its diffraction patterns and confirm that this phase is twice the atomic phase, analogous to frequency doubling in nonlinear optics. We also evaluate the parity and entanglement witnesses from the molecular diffraction patterns, from which we confirm the non-separability of the molecular state and characterize the nature of its entanglement.

Our experiment starts with a pure BEC of $1.2\times10^5$ cesium atoms with a temperature of $11(2)~\mathrm{nK}$ prepared in a two-dimensional disk-shaped, flat-bottomed trap with a radius of $21$~\textmu m. The atoms are strongly confined in the vertical direction to an optical dipole trap with trap frequency of $\omega_z=2\pi\times590~\mathrm{Hz}$. The condensate is prepared at a magnetic field of $B=20.25$~G with an atomic scattering length $a=180a_0$, where $a_0$ is the Bohr radius and the chemical potential is $\mu=k_B\times 40~\mathrm{nK}$~\cite{Zhang2021}. 

To synthesize molecules, the magnetic field is ramped across a $g-$wave Feshbach resonance at 19.849~G. During the ramp, the reaction occurs when the magnetic field is on resonance and terminates as the field moves away. The ramp converts the atomic BEC into a molecular BEC of up to $2\times10^4$ molecules with a temperature of $10(3)~\mathrm{nK}$~\cite{Zhang2021}. These molecules are spin-polarized and occupy a single high lying rovibrational state~\cite{Berninger2013}. The ramp is followed by a fast quench of the field to 17.2~G and then a resonant beam to remove any remaining atoms. To detect molecules, we dissociate them into atoms by ramping the magnetic field above the resonance and perform absorption imaging of the atoms.

In our first experiment, we perform matter-wave diffraction on the atomic and molecular BECs to verify their spatial coherence. A one-dimensional optical lattice with period $\lambda/2=532$~nm is applied on the sample. The lattice scatters particles to  modes with momentum $k=\pm2 k_L$, $\pm4 k_L$, ..., through two photon processes,  where $k_{L}=2\pi/\lambda$ is the lattice wavenumber, see Figs.~\ref{fig:2}a, c. The diffracted particles are spatially separated from the BEC after time-of-flight expansion~\cite{Supplement}. We realize Kapitza-Dirac diffraction with a static lattice potential, which off-resonantly scatters particles to multiple modes, see Figs.~\ref{fig:2}a, b, and Bragg diffraction with a moving lattice, which resonantly couples the BEC to the $|2k_L\rangle$ state, see Figs.~\ref{fig:2}c, d.

\begin{figure*}[ht!]
    \centering
    \includegraphics[width = \textwidth]{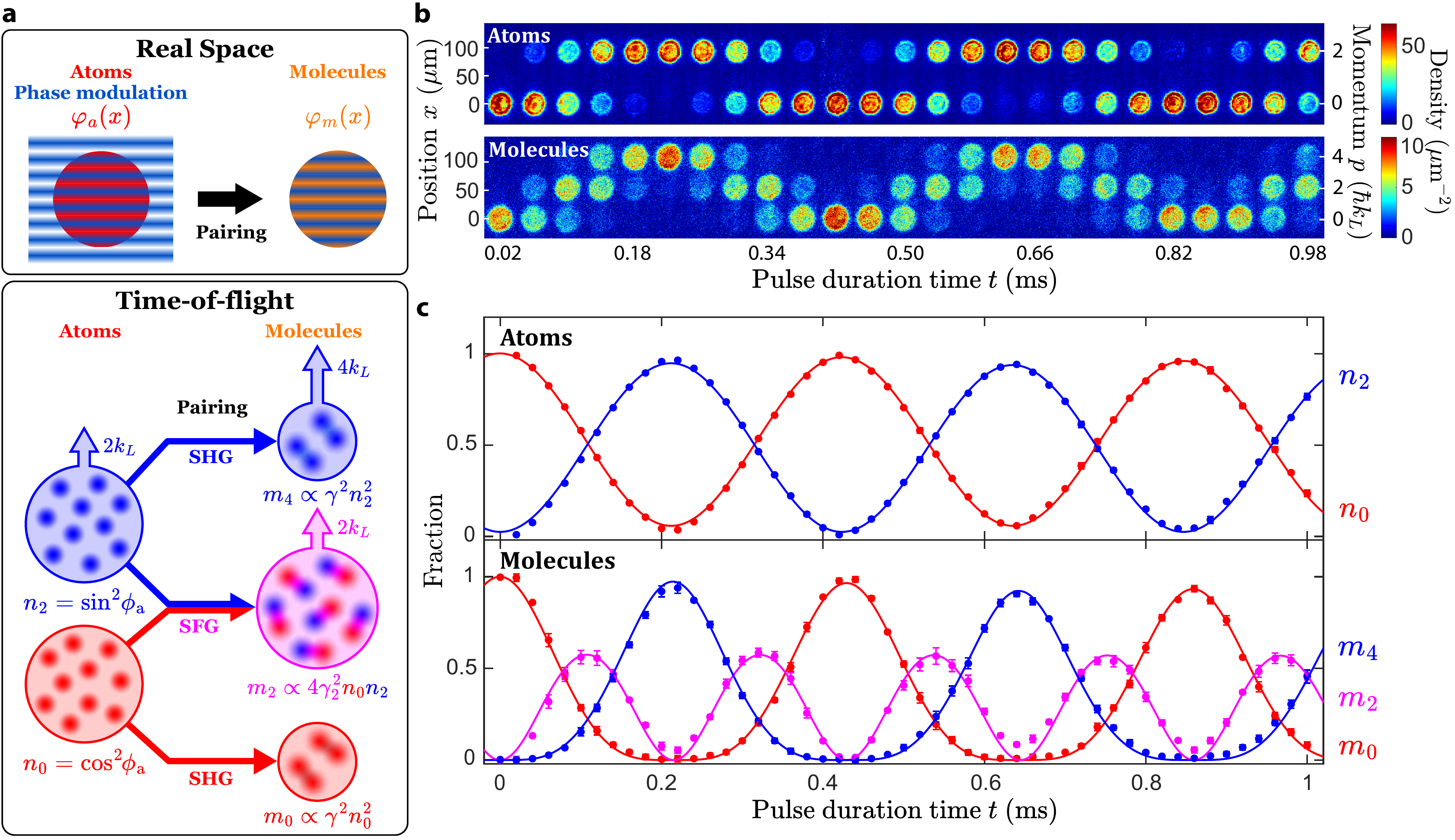}
    \caption{\textbf{Phase modulation on atomic and molecular BECs.} 
    \textbf{a}, Experimental protocol. A Bragg diffraction pulse imprints a phase $\varphi_\mathrm{a}(x)$ onto the atomic BEC. The atomic BEC is then converted into a molecular BEC with phase $\varphi_\mathrm{m}(x)$. %, which acquires a phase $\phi_\mathrm{m}(x)$. 
    The imprinted phase is reflected in the populations of diffracted atoms, $n_0=\cos^2\phi_\mathrm{a}$ and $n_2=\sin^2\phi_\mathrm{a}$. After pairing, molecules occupy momenta $k=0$, $2k_L$ and $4k_L$ which are proportional to the products of atomic populations. The molecular fractions of $m_0$ and $m_4$ are generated by second harmonic generation (SHG) with coupling constant $\gamma$ while $m_2$ is generated by sum frequency generation (SFG) with $\gamma_2$.
    The factor of four in $m_2$ comes from bosonic exchange symmetry. \textbf{b}, Oscillations of momentum state populations as a result of phase modulation on the atomic and molecular BECs. Images are taken after 16~ms time-of-flight expansion. Images shown here are averaged over five independent measurements. We note that the diffracted molecules move at half the speed of the atoms since they are twice as massive. \textbf{c}, Population fractions of diffracted atoms and molecules. Atomic fractions of $n_0$ and $n_2$ oscillate at a Rabi frequency of $\Omega_\mathrm{a}=2\pi\times 2.36(1)$~kHz. Molecular fractions are fit using $m_0=\cos^4\Omega_0t/2$, $m_2=A_2\sin^2\Omega_2t/2$ and $m_4=\sin^4\Omega_4t/2$, which give $\Omega_0=2\pi\times2.33(1)$, $\Omega_2=2\pi\times4.66(1)$, and $\Omega_4=2\pi\times2.33(1)$~kHz, respectively and $A_2=0.58(1)$. Error bars show 1-$\sigma$ standard deviations of the means.
    }
    \label{fig:Fig3}
\end{figure*}

Both diffraction processes induce coherent oscillations of atomic and molecular populations between zero and finite momentum modes. 
%4\pi\hbar k_L\tau_c/m
Remarkably, the low decoherence rate of the molecules $0.15~\mathrm{ms}^{-1}$ corresponds to a coherence length of $
\approx20$~\textmu m comparable with the size of the sample \cite{Supplement}. Therefore the spatial coherence is maintained during the pairing process. The oscillation frequencies of the diffracted populations also provide precise information on the energy dispersion of the particles in the lattice potential. By comparing the measurement results to the band theory calculation, we determine the molecular dynamical polarizability to be $\alpha_\mathrm{m}=k_B\times5.05(5)~\mathrm{nK}\cdot\mathrm{cm}^2/\mathrm{W}=1.95(2)\alpha_\mathrm{a}$, which is approximately twice the atomic polarizability $\alpha_\mathrm{a}$~\cite{Supplement}.
 
Bragg diffraction is a powerful tool to probe the phase relationship of atomic and molecular matter waves. A short Bragg pulse on the atomic BEC can imprint a phase pattern $\varphi_\mathrm{a}(x)$ on the wavefunction $\psi_\mathrm{a}(x)=\sqrt{\rho_\mathrm{a}(x)}e^{i\varphi_\mathrm{a}(x)}$. 
Following the pulse, we pair the atoms into molecules, and examine how the atomic phase transforms into phase $\varphi_\mathrm{m}(x)$ of the molecular wavefunction $\psi_\mathrm{m}(x)=\sqrt{\rho_\mathrm{m}(x)}e^{i\varphi_\mathrm{m}}(x)$. Here $\rho_\mathrm{a}$ and $\rho_\mathrm{m}$ are the atomic and molecular densities. The scheme is illustrated in Fig.~\ref{fig:Fig3}a.  

The phase information of the atomic and molecular matter waves can be readily extracted from their diffraction patterns. To see this, we express a wavefunction $\psi(x)$ as
\begin{align}
    \psi(x) = \sqrt{\rho(x)}e^{i\varphi(x)} 
    = \sum_{j=0, \pm1, \pm2...} \psi_{2j} e^{2i j k_L x},
    \label{eq3:wf_complex_number}
\end{align} 
where $\rho(x)$ is the particle density and $\psi_{2j}$ is the amplitude of the plane wave with momentum $2jk_L$. For Bragg-diffracted atoms, the single particle wavefunction $\psi_\mathrm{a}(x) =\cos(\Omega_\mathrm{a} t/2)-i\sin(\Omega_\mathrm{a} t/2)e^{2ik_Lx}$ can be modeled by a two level system undergoing Rabi oscillations with frequency $\Omega_\mathrm{a}$. 
We evaluate the phase modulation amplitude $\phi_a$ at the anti-nodes of the lattice $\phi_a\equiv\varphi_\mathrm{a}(x=\lambda/4)$, which evolves at a rate given by the Rabi frequency, $\phi_\mathrm{a}=\Omega_\mathrm{a} t/2$. The phase modulation amplitude can be extracted in our experiment from the atomic population fractions $n_0$ and $n_2$ based on

\begin{align}
    \cot^2\phi_\mathrm{a} = \frac{n_0}{n_2}.
    \label{eq4}
\end{align}

The pairing of atoms in two momentum modes results in molecules occupying three modes described by the wavefunction $\psi_\mathrm{m}(x)=\psi_\mathrm{m_0}+\psi_\mathrm{m_2}e^{2ik_L x}+ \psi_\mathrm{m_4}e^{4ik_L x}$. Molecules with momentum $k=0$ and $4 k_L$ come from pairing atoms with the same momentum $k=0$ or $2 k_L$, the matter-wave analogue of second harmonic generation (SHG). 
The molecular populations scale as $m_0\propto \gamma^2n_0^2$ and $m_4\propto \gamma^2n_2^2$, where $\gamma$ is the Feshbach coupling strength. 
Molecules with momentum $k=2 k_L$ come from pairing atoms with distinct momenta $k=0$ and $2k_L$, the matter-wave analogue of sum frequency generation (SFG). The $k=2 k_L$ population scales as $m_2\propto4\gamma_2^2 n_0 n_2$, where $\gamma_2$ is the sum frequency coupling strength and the factor of 4 comes from  bosonic symmetrization~\cite{Supplement}, see Fig.~\ref{fig:Fig3}a.

We measure the molecular populations after time-of-flight expansion. The normalized populations 
$m_0=|\psi_{\mathrm{m_0}}|^2$ and 
$m_4=|\psi_{\mathrm{m_4}}|^2$ 
oscillate at the atomic Rabi frequency $\Omega_\mathrm{a}$. They reach unity when atoms occupy a single momentum state after $0$, $\pi$, $2\pi$, $\ldots$ Bragg pulses. On the other hand, the population  
$m_2=|\psi_{\mathrm{m_2}}|^2$
oscillates at twice the Rabi frequency $2\Omega_\mathrm{a}$ and reaches the maximum $A_2=0.58(1)$ after $\pi/2$, $3\pi/2,\ldots$ Bragg pulses on the atoms, see Figs.~\ref{fig:Fig3}b and c. 

The model accurately describes the measurement results, and we obtain the ratio of the coupling constants $\gamma_2/\gamma=0.82(3)$. The smaller value of  $\gamma_2$ compared to $\gamma$ is consistent with our expectation that the sum frequency coupling is off-resonant by two atomic recoil energies, resulting in a lower coupling efficiency~\cite{Supplement}.

The molecular diffraction pattern reveals the phase modulation of the molecular matter waves. Applying Eq.~(\ref{eq3:wf_complex_number}), we evaluate the amplitude of the phase modulation on the molecules according to

\begin{align}
    \cot\phi_\mathrm{m} =  \frac{\sqrt{m_0}-\sqrt{m_4}}{\sqrt{m_2}}.
    \label{eq5}
\end{align}

\begin{figure}[t!]
    \centering
    \includegraphics[width = 0.5\textwidth]{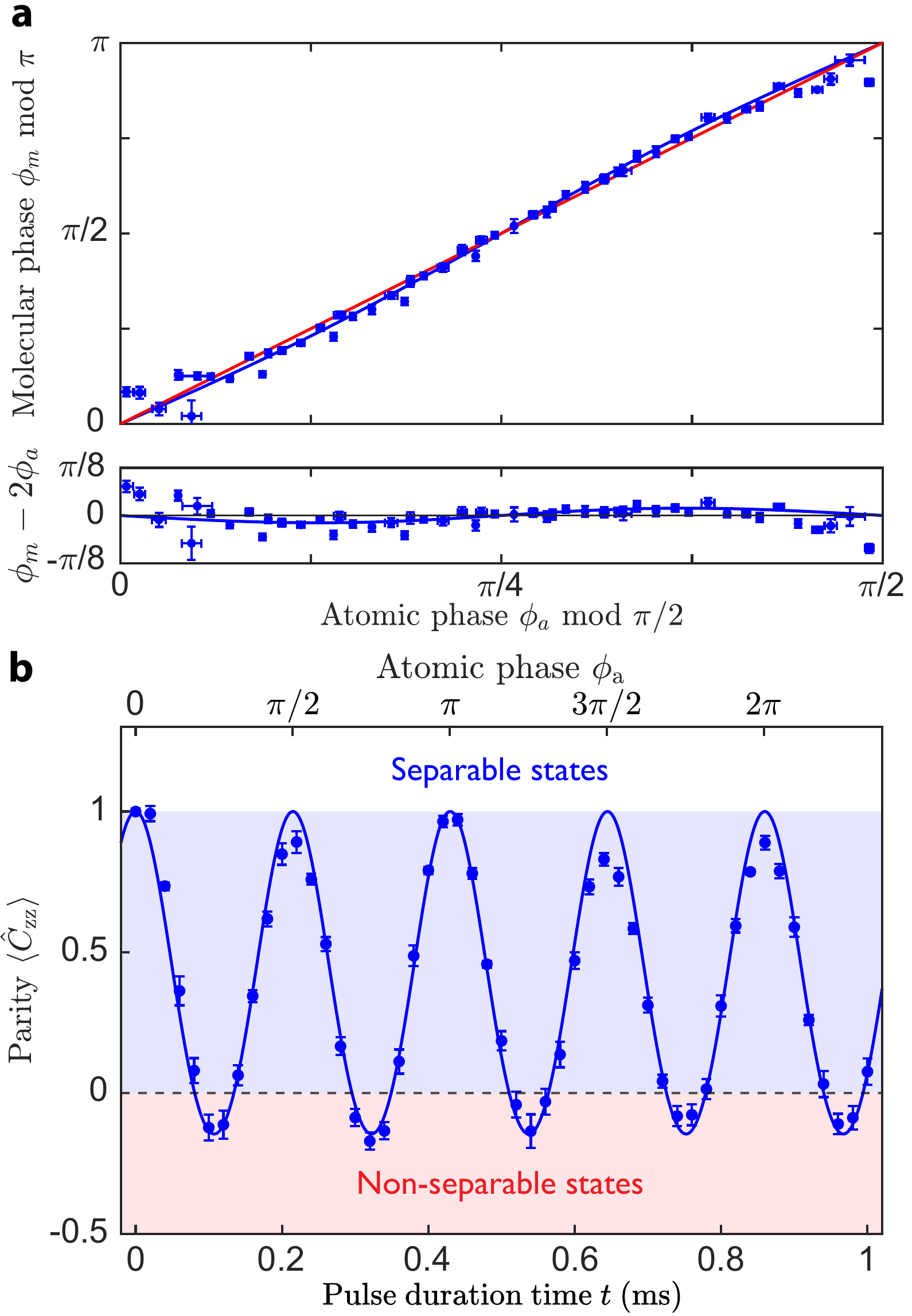}
    \caption{\textbf{Matter-wave phase doubling and entanglement generation.} 
    \textbf{a}, Dependence of the molecular phase $\phi_\mathrm{m}$  on the atomic phase $\phi_\mathrm{a}$ . The phases $\phi_\mathrm{a}$ and $\phi_\mathrm{m}$ are extracted from the diffraction patterns shown in Fig.~\ref{fig:Fig3}. The measurements (blue circles) are compared to the red line which represents a linear ratio of $\phi_\mathrm{m}/\phi_\mathrm{a}=2$. The blue line is the fit to the theoretical model $\phi_\mathrm{m}=2\phi_\mathrm{a}-\epsilon\sin4\phi_\mathrm{a}$. The lower panel shows the deviation of the data from the phase doubling ratio $\phi_\mathrm{m}=2\phi_\mathrm{a}$. The residual is fit with $-\epsilon\sin4\phi_\mathrm{a}$ (blue line), from which we obtain $\epsilon=0.06(2)$.
    \textbf{b}, Parity $\langle \hat{C}_{\mathrm{zz}} \rangle=m_0-m_2+m_4$ for molecules. Negative parity $-1 \leq \langle \hat{C}_{\mathrm{zz}} \rangle<0$ indicates non-separability of the two-atom momentum state. Error bars show 1-$\sigma$ standard deviation of the mean of 5 measurements.}
    \label{fig:Fig4}
\end{figure}

\noindent We compare the molecular phase modulation amplitude $\phi_\mathrm{m}\equiv\varphi_\mathrm{m}(x=\lambda/4)$ to the atomic phase modulation amplitude $\phi_\mathrm{a}$, see Fig.~\ref{fig:Fig4}a. The molecular phase evolves twice as much as the atomic phase $\phi_\mathrm{m}=2\phi_\mathrm{a}$ over several Rabi cycles, thus confirming phase doubling occurs during the synthesis of molecules. 

A closer examination suggests a small nonlinear correction to the exact phase doubling. The correction comes from the imbalance of the SFG and SHG coupling strengths $\gamma_2\neq \gamma$, and is predicted to be~\cite{Supplement}

\begin{equation}
    \phi_\mathrm{m}-2\phi_\mathrm{a}=-\epsilon\sin4\phi_\mathrm{a}+\mathcal{O}(\epsilon^2),
\end{equation}

\noindent where we obtain $\epsilon=0.06(2)$ from the fit to the data in Fig.~\ref{fig:Fig4}a, which is in agreement with the theoretical value of $\epsilon =\frac12 (1-\gamma_2/\gamma)\approx0.09(2)$~\cite{Supplement}. 

The phase doubling experiment also reveals entanglement generation during the formation of molecules. 
We start by testing the separability of the atom pairs in the reaction by evaluating the parity of their momentum populations $C_{zz} = \langle \hat{\sigma}_{1,z}\otimes\hat{\sigma}_{2,z}\rangle=m_0-m_2+m_4$, where $\hat{\sigma}_{j,z}$ is the Pauli-$Z$ operator on the $j$-th atom~\cite{Supplement,Sackett2000}. For product states of two identical particles, the parity is constrained within $0\leq C_{zz}\leq 1$. Negative parity $-1\leq C_{zz}<0$ indicates the non-separability and entanglement of the atom pairs.

In Fig.~\ref{fig:Fig4}b, we show that the parity $C_{zz}$ oscillates at twice the atomic Rabi frequency. The parity reaches the maximum of $C_{zz}=1$, when the molecules are in a single momentum state $|0\rangle$ or $|4k_L\rangle$. The parity reaches the minimum of $C_{zz}=-0.15(1)<0$ and enters the non-separable regime, when atoms are prepared in an equal superposition of momentum states $|0\rangle$ and $|2k_L\rangle$ after a $\pi/2$, $3\pi/2$,... Bragg pulse. Here reacting atoms exhibit the highest degree of momentum entanglement.

To elucidate the nature of the entanglement, we perform quantum state tomography by evaluating the Bell state entanglement witnesses $W_{\Psi^{\pm}}$ and $W_{\Phi^{\pm}}$~\cite{Supplement,Hyllus2005}. These witnesses, evaluated from the momentum populations of molecules, quantify the overlap of the momentum states of atoms during the reaction with the Bell states. The largest positive value $W_{\Psi^+} = 0.29(2)$ is obtained after a $\pi/2$ pulse~\cite{Supplement}. This indicates that the momentum correlations are non-classical and the atomic wavefunction most closely resembles the Bell state $|\Psi^+\rangle =1/\sqrt{2}\left(|k=0,2k_\mathrm{L}\rangle+|2k_\mathrm{L},k=0\rangle\right)$. %Full reconstruction yields $\ket{\psi_\mathrm{m}} = \frac{1}{2}\sqrt{2-W_{\Psi^+}}\ket{\Phi^-} - \frac{i}{2}\sqrt{2+W_{\Psi^+}}\ket{\Psi^+}$~\cite{Supplement}.

We attribute the entanglement formation in our system to nonlinear wave mixing and bosonic statistics of the particles. The reaction Hamiltonian $\hat{H}\propto (\hat{\psi}_\mathrm{a_0}+\hat{\psi}_\mathrm{a_2})^2$, see Eq.~\eqref{eq2:atom_hamiltonian}, mixes the atomic matter waves, where $\hat{\psi}_{\mathrm{a}_j}$ is the field operator for the mode with momentum $j k_L$. The square produces the cross terms $\hat{\psi}_\mathrm{a_0}\hat{\psi}_\mathrm{a_2} + \hat{\psi}_\mathrm{a_2}\hat{\psi}_\mathrm{a_0}=2\hat{\psi}_\mathrm{a_2}\hat{\psi}_\mathrm{a_0}$ which add constructively due to bosonic symmetry. This results in an enhancement of the molecular population with momentum $k=2k_{\mathrm{L}}$ by a factor of 2, $m_2 \propto 4n_0n_2$, compared to the classical expectation of $2n_0n_2$. The enhancement of $m_2$ explains the reaction wavefunction approaching the Bell state $|\Psi^+\rangle$. 

In summary, we demonstrate phase-matched chemical reactions between atomic and molecular BECs at a Feshbach resonance. Using matter-wave diffraction as a probe, we observe phase doubling and entanglement generation emerging as hallmarks of nonlinear matter-wave mixing. These results establish a new regime of quantum chemistry where macroscopic quantum fields govern the reaction dynamics. Our phase imprinting scheme also lays the foundation to measure and track the reaction phase with the goal of controlling the  reaction dynamics. In addition, our work may pave the way for harnessing reactions to entangle atomic and molecular qubits for quantum information science. 

\noindent\textbf{Acknowledgement}
\noindent We thank S.~Cornish, X.~Lai, and S.~Mystakidis for helpful discussions. We thank Z.~Yan and J.~Jachinowski for carefully reading the manuscript. This work was supported by
the National Science Foundation under Grant No. PHY-1511696 and PHY-2103542, by the Air Force Office of Scientific Research under award number FA9550-21-1-0447. S.~N. acknowledges support from the Takenaka Scholarship Foundation.

\bibliographystyle{naturemag}
\bibliography{filteredbibliography}

@article{Wolf2017,
author = {Joschka Wolf  and Markus Deiß  and Artjom Krükow  and Eberhard Tiemann  and Brandon P. Ruzic  and Yujun Wang  and José P. D’Incao  and Paul S. Julienne  and Johannes Hecker Denschlag },
title = {State-to-state chemistry for three-body recombination in an ultracold rubidium gas},
journal = {Science},
volume = {358},
number = {6365},
pages = {921-924},
year = {2017},
doi = {10.1126/science.aan8721}
}

@article{Moore2002,
  title = {{B}ose-{E}nhanced chemistry: {A}mplification of selectivity in the dissociation of molecular {B}ose-{E}instein condensates},
  author = {Moore, M. G. and Vardi, A.},
  journal = {Phys. Rev. Lett.},
  volume = {88},
  issue = {16},
  pages = {160402},
  numpages = {4},
  year = {2002},
  month = {Apr},
  publisher = {American Physical Society},
  doi = {},
  url = {}
}

@article{Park2017,
author = {Jee Woo Park  and Zoe Z. Yan  and Huanqian Loh  and Sebastian A. Will  and Martin W. Zwierlein},
title = {{S}econd-scale nuclear spin coherence time of ultracold $^{23}$\uppercase{N}a$^{40}$\uppercase{K} molecules},
journal = {Science},
volume = {357},
number = {6349},
pages = {372-375},
year = {2017},
doi = {},
URL = {},
eprint = {}}

@article{RafacTanner1998,
  title = {{M}easurement of the ratio of the cesium ${D}$-line transition strengths},
  author = {Rafac, Robert J. and Tanner, Carol E.},
  journal = {Phys. Rev. A},
  volume = {58},
  issue = {2},
  pages = {1087--1097},
  numpages = {0},
  year = {1998},
  month = {Aug},
  publisher = {American Physical Society},
  doi = {},
  url = {}
}

@incollection{GRIMM200095,
title = {{O}ptical dipole traps for neutral atoms},
editor = {Benjamin Bederson and Herbert Walther},
series = {Adv. Atom. Mol. Opt. Phy.},
publisher = {Academic Press},
volume = {42},
pages = {95-170},
year = {2000},
issn = {1049-250X},
doi = {},
url = {},
author = {Rudolf Grimm and Matthias Weidemüller and Yurii B. Ovchinnikov}
}

@article{Schmiedmayer,
  title = {{M}atter-wave interferometers with trapped strongly interacting {F}eshbach molecules},
  author = {Li, Chen and Liang, Qi and Paranjape, Pradyumna and Wu, RuGway and Schmiedmayer, J\"org},
  journal = {Phys. Rev. Res.},
  volume = {6},
  issue = {2},
  pages = {023217},
  numpages = {12},
  year = {2024},
  month = {May},
  publisher = {American Physical Society},
  doi = {},
  url = {}
}

@misc{Supplement,
  note = {See supplementary materials}
}

@Article{Bongs2019,
  author    = {Bongs, Kai and Holynski, Michael and Vovrosh, Jamie and Bouyer, Philippe and Condon, Gabriel and Rasel, Ernst and Schubert, Christian and Schleich, Wolfgang P. and Roura, Albert},
  journal   = {Nat. Rev. Phys.},
  title     = {Taking atom interferometric quantum sensors from the laboratory to real-world applications},
  year      = {2019},
  issn      = {2522-5820},
  month     = oct,
  number    = {12},
  pages     = {731--739},
  volume    = {1},
  doi       = {10.1038/s42254-019-0117-4},
  publisher = {Springer Science and Business Media LLC},
}

@Article{Stray2022,
  author    = {Stray, Ben and Lamb, Andrew and Kaushik, Aisha and Vovrosh, Jamie and Rodgers, Anthony and Winch, Jonathan and Hayati, Farzad and Boddice, Daniel and Stabrawa, Artur and Niggebaum, Alexander and Langlois, Mehdi and Lien, Yu-Hung and Lellouch, Samuel and Roshanmanesh, Sanaz and Ridley, Kevin and de Villiers, Geoffrey and Brown, Gareth and Cross, Trevor and Tuckwell, George and Faramarzi, Asaad and Metje, Nicole and Bongs, Kai and Holynski, Michael},
  journal   = {Nature},
  title     = {Quantum sensing for gravity cartography},
  year      = {2022},
  issn      = {1476-4687},
  month     = feb,
  number    = {7898},
  pages     = {590--594},
  volume    = {602},
  doi       = {10.1038/s41586-021-04315-3},
  publisher = {Springer Science and Business Media LLC},
}

@Article{Brand2020,
  author    = {Brand, Christian and Kiałka, Filip and Troyer, Stephan and Knobloch, Christian and Simonović, Ksenija and Stickler, Benjamin A. and Hornberger, Klaus and Arndt, Markus},
  journal   = {Phys. Rev. Lett.},
  title     = {Bragg {D}iffraction of {L}arge {O}rganic {M}olecules},
  year      = {2020},
  issn      = {1079-7114},
  month     = jul,
  number    = {3},
  pages     = {033604},
  volume    = {125},
  doi       = {10.1103/physrevlett.125.033604},
  publisher = {American Physical Society (APS)},
}

@Article{Bohn2017,
  author    = {Bohn, John L. and Rey, Ana Maria and Ye, Jun},
  journal   = {Science},
  title     = {Cold molecules: {P}rogress in quantum engineering of chemistry and quantum matter},
  year      = {2017},
  issn      = {1095-9203},
  month     = sep,
  number    = {6355},
  pages     = {1002--1010},
  volume    = {357},
  doi       = {10.1126/science.aam6299},
  publisher = {American Association for the Advancement of Science (AAAS)},
}

@Article{Heinzen2000,
  author    = {Heinzen, D. J. and Wynar, Roahn and Drummond, P. D. and Kheruntsyan, K. V.},
  journal   = {Phys. Rev. Lett.},
  title     = {Superchemistry: {D}ynamics of coupled atomic and molecular \uppercase{B}ose–\uppercase{E}instein condensates},
  year      = {2000},
  issn      = {1079-7114},
  month     = may,
  number    = {22},
  pages     = {5029--5033},
  volume    = {84},
  doi       = {10.1103/physrevlett.84.5029},
  publisher = {American Physical Society (APS)},
}

@Article{Chen2024,
  author    = {Chen, Xing-Yan and Biswas, Shrestha and Eppelt, Sebastian and Schindewolf, Andreas and Deng, Fulin and Shi, Tao and Yi, Su and Hilker, Timon A. and Bloch, Immanuel and Luo, Xin-Yu},
  journal   = {Nature},
  title     = {Ultracold field-linked tetratomic molecules},
  year      = {2024},
  issn      = {1476-4687},
  month     = jan,
  number    = {7998},
  pages     = {283--287},
  volume    = {626},
  doi       = {10.1038/s41586-023-06986-6},
  publisher = {Springer Science and Business Media LLC},
}

@Article{Broglie1925,
  author    = {de Broglie, Louis},
  journal   = {Ann. Phys.-Paris},
  title     = {Recherches sur la théorie des quanta},
  year      = {1925},
  issn      = {1286-4838},
  number    = {3},
  pages     = {22--128},
  volume    = {10},
  doi       = {10.1051/anphys/192510030022},
  publisher = {EDP Sciences},
}

@Article{Davisson1927,
  author    = {Davisson, C. and Germer, L. H.},
  journal   = {Phys. Rev.},
  title     = {Diffraction of {E}lectrons by a {C}rystal of {N}ickel},
  year      = {1927},
  issn      = {0031-899X},
  month     = dec,
  number    = {6},
  pages     = {705--740},
  volume    = {30},
  doi       = {10.1103/physrev.30.705},
  publisher = {American Physical Society (APS)},
}

@Article{Anwar2021,
  author    = {Anwar, Ali and Perumangatt, Chithrabhanu and Steinlechner, Fabian and Jennewein, Thomas and Ling, Alexander},
  journal   = {Rev. Sci. Instrum.},
  title     = {Entangled photon-pair sources based on three-wave mixing in bulk crystals},
  year      = {2021},
  issn      = {1089-7623},
  month     = apr,
  number    = {4},
  volume    = {92},
  doi       = {10.1063/5.0023103},
  publisher = {AIP Publishing},
}

@Article{Estermann1930,
  author    = {Estermann, I. and Stern, O.},
  journal   = {Z. Phys.},
  title     = {Beugung von {M}olekularstrahlen},
  year      = {1930},
  issn      = {1434-601X},
  month     = jan,
  number    = {1–2},
  pages     = {95--125},
  volume    = {61},
  doi       = {10.1007/bf01340293},
  publisher = {Springer Science and Business Media LLC},
}

@Article{Greenberger1983,
  author    = {Greenberger, Daniel M.},
  journal   = {Rev. Mod. Phys.},
  title     = {The neutron interferometer as a device for illustrating the strange behavior of quantum systems},
  year      = {1983},
  issn      = {0034-6861},
  month     = oct,
  number    = {4},
  pages     = {875--905},
  volume    = {55},
  doi       = {10.1103/revmodphys.55.875},
  publisher = {American Physical Society (APS)},
}

@Article{Malla2022,
  author    = {Malla, Rajesh K. and Chernyak, Vladimir Y. and Sun, Chen and Sinitsyn, Nikolai A.},
  journal   = {Phys. Rev. Lett.},
  title     = {Coherent reaction between molecular and atomic \uppercase{B}ose–\uppercase{E}instein condensates: integrable model},
  year      = {2022},
  issn      = {1079-7114},
  month     = jul,
  number    = {3},
  pages     = {033201},
  volume    = {129},
  doi       = {10.1103/physrevlett.129.033201},
  publisher = {American Physical Society (APS)},
}

@Article{Burchesky2021,
  author    = {Burchesky, Sean and Anderegg, Loïc and Bao, Yicheng and Yu, Scarlett S. and Chae, Eunmi and Ketterle, Wolfgang and Ni, Kang-Kuen and Doyle, John M.},
  journal   = {Phys. Rev. Lett.},
  title     = {Rotational coherence times of polar molecules in optical tweezers},
  year      = {2021},
  issn      = {1079-7114},
  month     = sep,
  number    = {12},
  pages     = {123202},
  volume    = {127},
  doi       = {10.1103/physrevlett.127.123202},
  publisher = {American Physical Society (APS)},
}

@Article{Langen2024,
  author    = {Langen, Tim and Valtolina, Giacomo and Wang, Dajun and Ye, Jun},
  journal   = {Nat. Phys.},
  title     = {Quantum state manipulation and cooling of ultracold molecules},
  year      = {2024},
  issn      = {1745-2481},
  month     = may,
  number    = {5},
  pages     = {702--712},
  volume    = {20},
  doi       = {10.1038/s41567-024-02423-1},
  publisher = {Springer Science and Business Media LLC},
}

@Article{Zhang2021,
  author    = {Zhang, Zhendong and Chen, Liangchao and Yao, Kai-Xuan and Chin, Cheng},
  journal   = {Nature},
  title     = {Transition from an atomic to a molecular \uppercase{B}ose–\uppercase{E}instein condensate},
  year      = {2021},
  issn      = {1476-4687},
  month     = apr,
  number    = {7856},
  pages     = {708--711},
  volume    = {592},
  doi       = {10.1038/s41586-021-03443-0},
  publisher = {Springer Science and Business Media LLC},
}

@Article{Bigagli2024,
  author    = {Bigagli, Niccolò and Yuan, Weijun and Zhang, Siwei and Bulatovic, Boris and Karman, Tijs and Stevenson, Ian and Will, Sebastian},
  journal   = {Nature},
  title     = {Observation of \uppercase{B}ose–\uppercase{E}instein condensation of dipolar molecules},
  year      = {2024},
  issn      = {1476-4687},
  month     = jun,
  number    = {8020},
  pages     = {289--293},
  volume    = {631},
  doi       = {10.1038/s41586-024-07492-z},
  publisher = {Springer Science and Business Media LLC},
}

@Article{Ruttley2025,
  author    = {Ruttley, Daniel K. and Hepworth, Tom R. and Guttridge, Alexander and Cornish, Simon L.},
  journal   = {Nature},
  title     = {Long-lived entanglement of molecules in magic-wavelength optical tweezers},
  year      = {2025},
  issn      = {1476-4687},
  month     = jan,
  number    = {8047},
  pages     = {827--832},
  volume    = {637},
  doi       = {10.1038/s41586-024-08365-1},
  publisher = {Springer Science and Business Media LLC},
}

@Article{Chin2010,
  author    = {Chin, Cheng and Grimm, Rudolf and Julienne, Paul and Tiesinga, Eite},
  journal   = {Rev. Mod. Phys.},
  title     = {Feshbach resonances in ultracold gases},
  year      = {2010},
  issn      = {1539-0756},
  month     = apr,
  number    = {2},
  pages     = {1225--1286},
  volume    = {82},
  doi       = {10.1103/revmodphys.82.1225},
  publisher = {American Physical Society (APS)},
}

@Article{DellAnno2006,
  author    = {Dell’Anno, Fabio and De Siena, Silvio and Illuminati, Fabrizio},
  journal   = {Phys. Rep.},
  title     = {Multiphoton quantum optics and quantum state engineering},
  year      = {2006},
  issn      = {0370-1573},
  month     = may,
  number    = {2–3},
  pages     = {53--168},
  volume    = {428},
  doi       = {10.1016/j.physrep.2006.01.004},
  publisher = {Elsevier BV},
}

@Article{Adesso2007,
  author    = {Adesso, Gerardo and Illuminati, Fabrizio},
  journal   = {J. Phys. A-Math. Theor.},
  title     = {Entanglement in continuous-variable systems: recent advances and current perspectives},
  year      = {2007},
  issn      = {1751-8121},
  month     = jun,
  number    = {28},
  pages     = {7821--7880},
  volume    = {40},
  doi       = {10.1088/1751-8113/40/28/s01},
  publisher = {IOP Publishing},
}

@Article{Cornish2024,
  author    = {Cornish, Simon L. and Tarbutt, Michael R. and Hazzard, Kaden R. A.},
  journal   = {Nat. Phys.},
  title     = {Quantum computation and quantum simulation with ultracold molecules},
  year      = {2024},
  issn      = {1745-2481},
  month     = may,
  number    = {5},
  pages     = {730--740},
  volume    = {20},
  doi       = {10.1038/s41567-024-02453-9},
  publisher = {Springer Science and Business Media LLC},
}

@Article{Bao2023,
  author    = {Bao, Yicheng and Yu, Scarlett S. and Anderegg, Loïc and Chae, Eunmi and Ketterle, Wolfgang and Ni, Kang-Kuen and Doyle, John M.},
  journal   = {Science},
  title     = {Dipolar spin-exchange and entanglement between molecules in an optical tweezer array},
  year      = {2023},
  issn      = {1095-9203},
  month     = dec,
  number    = {6675},
  pages     = {1138--1143},
  volume    = {382},
  doi       = {10.1126/science.adf8999},
  publisher = {American Association for the Advancement of Science (AAAS)},
}

@Article{Martins2025,
  author    = {Martins, Fernanda B. V. and Schmutz, Hansjürg and Agner, Josef A. and Zhelyazkova, Valentina and Merkt, Frédéric},
  journal   = {Phys. Rev. Lett.},
  title     = {Microwave-controlled cold chemistry},
  year      = {2025},
  issn      = {1079-7114},
  month     = mar,
  number    = {12},
  pages     = {123401},
  volume    = {134},
  doi       = {10.1103/physrevlett.134.123401},
  publisher = {American Physical Society (APS)},
}

@Article{Holland2023,
  author    = {Holland, Connor M. and Lu, Yukai and Cheuk, Lawrence W.},
  journal   = {Science},
  title     = {On-demand entanglement of molecules in a reconfigurable optical tweezer array},
  year      = {2023},
  issn      = {1095-9203},
  month     = dec,
  number    = {6675},
  pages     = {1143--1147},
  volume    = {382},
  doi       = {10.1126/science.adf4272},
  publisher = {American Association for the Advancement of Science (AAAS)},
}

@Article{Zhang2023,
author={Zhang, Zhendong
and Nagata, Shu
and Yao, Kai-Xuan
and Chin, Cheng},
title={Many-body chemical reactions in a quantum degenerate gas},
journal={Nat. Phys.},
year={2023},
month={Oct},
day={01},
volume={19},
number={10},
pages={1466-1470},
issn={1745-2481},
doi={10.1038/s41567-023-02139-8}
}

@Article{Poulsen2001,
  author    = {Poulsen, Uffe V. and Mølmer, Klaus},
  journal   = {Phys. Rev. A},
  title     = {Quantum states of {B}ose-{E}instein condensates formed by molecular dissociation},
  year      = {2001},
  issn      = {1094-1622},
  month     = jan,
  number    = {2},
  pages     = {023604},
  volume    = {63},
  doi       = {10.1103/physreva.63.023604},
  publisher = {American Physical Society (APS)},
}

@Article{Sackett2000,
author={Sackett, C. A.
and Kielpinski, D.
and King, B. E.
and Langer, C.
and Meyer, V.
and Myatt, C. J.
and Rowe, M.
and Turchette, Q. A.
and Itano, W. M.
and Wineland, D. J.
and Monroe, C.},
title={Experimental entanglement of four particles},
journal={Nature},
year={2000},
month={Mar},
day={01},
volume={404},
number={6775},
pages={256-259},
issn={1476-4687},
doi={10.1038/35005011}
}

@Article{Agarwal1969,
  author    = {Agarwal, G.S.},
  journal   = {Opt. Commun.},
  title     = {Quantum theory of second harmonic generation},
  year      = {1969},
  issn      = {0030-4018},
  month     = jul,
  number    = {3},
  pages     = {132--134},
  volume    = {1},
  doi       = {10.1016/0030-4018(69)90029-7},
  publisher = {Elsevier BV},
}

@Article{Hyllus2005,
  author    = {Hyllus, Philipp and Gühne, Otfried and Bruß, Dagmar and Lewenstein, Maciej},
  journal   = {Phys. Rev. A},
  title     = {Relations between entanglement witnesses and Bell inequalities},
  year      = {2005},
  issn      = {1094-1622},
  month     = jul,
  number    = {1},
  pages     = {012321},
  volume    = {72},
  doi       = {10.1103/physreva.72.012321},
  publisher = {American Physical Society (APS)},
}

@Article{Liu2024,
  author    = {Liu, Yi-Xiang and Zhu, Lingbang and Luke, Jeshurun and Houwman, J. J. Arfor and Babin, Mark C. and Hu, Ming-Guang and Ni, Kang-Kuen},
  journal   = {Science},
  title     = {Quantum interference in atom-exchange reactions},
  year      = {2024},
  issn      = {1095-9203},
  month     = jun,
  number    = {6700},
  pages     = {1117--1121},
  volume    = {384},
  doi       = {10.1126/science.adl6570},
  publisher = {American Association for the Advancement of Science (AAAS)},
}

@Article{Croft2017,
  author    = {Croft, J. F. E. and Makrides, C. and Li, M. and Petrov, A. and Kendrick, B. K. and Balakrishnan, N. and Kotochigova, S.},
  journal   = {Nat. Commun.},
  title     = {Universality and chaoticity in ultracold {K}+{KR}b chemical reactions},
  year      = {2017},
  issn      = {2041-1723},
  month     = jul,
  number    = {1},
  volume    = {8},
  doi       = {10.1038/ncomms15897},
  publisher = {Springer Science and Business Media LLC},
}

@article{Zhou2020,
  title = {Second-Scale Coherence Measured at the Quantum Projection Noise Limit with Hundreds of Molecular Ions},
  author = {Zhou, Yan and Shagam, Yuval and Cairncross, William B. and Ng, Kia Boon and Roussy, Tanya S. and Grogan, Tanner and Boyce, Kevin and Vigil, Antonio and Pettine, Madeline and Zelevinsky, Tanya and Ye, Jun and Cornell, Eric A.},
  journal = {Phys. Rev. Lett.},
  volume = {124},
  issue = {5},
  pages = {053201},
  numpages = {5},
  year = {2020},
  month = {Feb},
  publisher = {American Physical Society},
  doi = {},
  url = {}
}

@article{Berninger2013,
  title = {Feshbach resonances, weakly bound molecular states, and coupled-channel potentials for cesium at high magnetic fields},
  author = {Berninger, Martin and Zenesini, Alessandro and Huang, Bo and Harm, Walter and N\"agerl, Hanns-Christoph and Ferlaino, Francesca and Grimm, Rudolf and Julienne, Paul S. and Hutson, Jeremy M.},
  journal = {Phys. Rev. A},
  volume = {87},
  issue = {3},
  pages = {032517},
  numpages = {17},
  year = {2013},
  month = {Mar},
  publisher = {American Physical Society}
}

@article{Gong2003,
    author = {Gong, Jiangbin and Shapiro, Moshe and Brumer, Paul},
    title = {Entanglement-assisted coherent control in nonreactive diatom–diatom scattering},
    journal = {The Journal of Chemical Physics},
    volume = {118},
    number = {6},
    pages = {2626-2636},
    year = {2003},
    month = {02},
    issn = {0021-9606},
    doi = {10.1063/1.1535428}
}

@article{Li2019,
author = {Junxu Li  and Sabre Kais },
title = {Entanglement classifier in chemical reactions},
journal = {Science Advances},
volume = {5},
number = {8},
pages = {eaax5283},
year = {2019},
doi = {10.1126/sciadv.aax5283}
}

\noindent\textbf{Author contributions}

S.N. performed the experiments. 
S.N., T.M. and C.K. analysed the data. 
All authors contributed to the discussion and interpretation of the data.
C.C. developed the theory.
S.N., T.M. and C.C. wrote the manuscript.
C.C. supervised the project.

\noindent\textbf{Competing interests}

The authors declare no competing financial interests.

\noindent\textbf{Materials and correspondence}

Correspondence and reasonable requests for materials and data should be addressed to C.~C. (cchin@uchicago.edu).

\noindent\textbf{Data availability}

The data that support the plots within this paper and other findings of this study are available from the corresponding author upon reasonable request.

\noindent\textbf{Code availability}

The codes for the analysis of data shown within this paper are available from the corresponding author upon reasonable request.

\end{document}

%\linenumbers
\widetext
\setcounter{page}{15}
\setcounter{equation}{0}
\setcounter{figure}{0}
\setcounter{table}{0}
\makeatletter
\renewcommand{\theequation}{M\arabic{equation}}
\renewcommand{\thefigure}{M\arabic{figure}}
\renewcommand{\thetable}{M\arabic{table}}

\begin{center}
\begin{Large}
Methods for\\
\end{Large}
\begin{large}
\textbf{Observation of Phase Doubling and Entanglement in Coherent
Matter-Wave Reactions}
\end{large}\\
Shu Nagata,$^{1}$ Tadej Me{\v z}nar{\v s}i{\v c},$^{1}$ Chuixin Kong,$^{1}$ and Cheng Chin$^{*1}$\\
\textit{$^{1}$ The James Franck Institute, Enrico Fermi Institute and Department of Physics, University of Chicago, Chicago, IL 60637, USA}\\
\end{center}

\section{I. Preparation of atomic and molecular BEC\lowercase{s}}
Our experiment starts with $1.2\times10^5$ Bose-condensed $^{\mathrm{133}}\mathrm{Cs}$ atoms in a two-dimensional (2D) cylindrical flat-bottomed trap at magnetic field $B\approx20.3$~G. The atoms are polarized in the ground state $|F=3,m_F=3\rangle$ where $F$ and $m_F$ are the quantum numbers for the total angular momentum and its projection along the direction of the magnetic field, respectively. 
The atoms are trapped by a blue-detuned ring-like barrier generated by a digital micromirror device (DMD) in the radial direction and an optical dipole trap with a trap frequency of $\omega_z=2\pi\times 590~\mathrm{Hz}$ in the vertical direction. To create molecules, we rapidly sweep the magnetic field across the narrow $g-$wave Feshbach resonance at $19.849$~G with a width of $8.3$~mG in $1.5$~ms to couple the Cs atoms into $\mathrm{Cs}_2$ molecules in $|f=4,m_f=4;l=4,m_l=2\rangle$ where $f$ is the total angular momentum quantum number of the molecule and $l$ is the molecular orbital angular momentum quantum number, and $m_f$ and $m_l$ are the projections of $f$ and $l$ along the direction of the magnetic field, respectively~\cite{Zhang2021,Berninger2013}. After the sweep, we jump the magnetic field far away from the resonance to $17.2$~G and blow away any remaining atoms using a resonant light pulse. To image the molecules, we dissociate them back into atoms by sweeping across the Feshbach resonance again from $19.5$ to $20.4$~G in less than $1$~ms and image the atoms using absorption imaging.

\section{II. Optical lattice diffraction experiment}
To diffract the particles, we apply a pulse of an optical standing wave from a Nd:YAG laser with wavelength $\lambda=1064$~nm. The standing wave is prepared by retro-reflecting the incident beam which passes through the sample and then a pair of acousto-optic modulators (AOMs) which we use to tune the frequency of the retro-reflected beam. For Kapitza-Dirac diffraction, we set the retro-reflected beam to have the same frequency as the incident beam to create a static lattice.

For Bragg diffraction, the retro-reflected beam is detuned by $\Delta f=5.304~\mathrm{kHz}$ and $\Delta f=2.652~\mathrm{kHz}$ for atoms and molecules, respectively, to resonantly transfer particles to momentum $k=2 k_L$. This is realized by setting the AOMs to $+80~\mathrm{MHz}+\Delta f/2$ and $-80~\mathrm{MHz}$. The retro-reflected beam passes through the AOMs twice and acquires the desired detuning $\Delta f$.

To measure the atomic and molecular populations in different momentum modes, we perform a 2D time-of-flight (TOF) expansion to spatially separate atoms and molecules in different orders. We expand the sample in the radial $x-y$ direction by shutting off the DMD potential barrier while keeping the dipole trap turned on in the $z-$direction such that the sample only expands radially. Before the lattice pulse, we quench the scattering length to $0a_0$ for atoms. The scattering length of the molecules is fixed at $220a_0$~\cite{Zhang2021}.

After $t_\mathrm{TOF}=$~15~ms, when the different diffracted orders are sufficiently separated, we image the sample and determine the populations by counting the particle number in each momentum mode. The TOF is short enough such that the diffracted components maintain the disk-shape of the initial BEC. To obtain the precise particle number, we carefully evaluate and compensate for the background signal and integrate the particle numbers within the circular area of the BEC for each diffracted mode. This procedure allows us to determine the atomic and molecular populations shown in Figs.$2$ and $3$.

We use an empirical function to fit the atomic and molecular Rabi oscillations in Fig.~$2$
\begin{equation}
    P(t) = P(0)\cos(\Omega t+\theta)e^{-\eta t}+C,
\end{equation}
where $P$ is the population, $\theta$ is the phase shift, $\eta$ is the decoherence rate and $C$ is a constant. The results from the fitting are listed in Tables~\ref{Table：S1} and \ref{Table：S2}. From these measurement results, we determine the lattice potential depth to be $4.70(2)$ and $9.31(2)~E_\mathrm{R}$ for atoms and molecules, respectively, in Fig.~$2$b for Kaptiza-Dirac diffraction, and $2.41(1)$ and $4.83(1)~E_\mathrm{R}$ for atoms and molecules, respectively, in Fig.~$2$d for Bragg diffraction. Since the lattice beam intensity is the same for both atoms and molecules, the different potential depths for atoms and molecules suggest a difference in their dynamical polarizabilities.

We estimate the coherence length as $l_c=v_\mathrm{rel}\tau_c$ where $v_\mathrm{rel}=2\pi\hbar/\lambda m$ is the relative velocity of the excited mode and $\tau_c=1/\eta$ is the coherence time. We measure a decoherence rate of $\eta=0.15\mathrm{ms}^{-1}$ from the Bragg diffraction of molecules. This gives us a coherence length of $l_c\approx 20$~\textmu m which is comparable to the size of our sample.

\begin{table}[h!]
    \centering
    \caption{\textbf{Oscillation frequencies and decoherence rates of the Kapitza-Dirac diffracted atomic and molecular populations.} Obtained from the fits in Fig.~$2$b.}
    \begin{ruledtabular}
    \begin{tabular}{c c c c c c}
    \multicolumn{3}{c}{\textbf{Atoms}} & \multicolumn{3}{c}{\textbf{Molecules}}\\
    \hline
    Momentum & Frequency & Decoherence rate & Momentum & Frequency & Deoherence rate\\
    $k$ & $\Omega/2\pi$~(kHz) & $\eta$~$\left(\mathrm{ms}^{-1}\right)$ & $k$ & $\Omega/2\pi$~(kHz) & $\eta$~$\left(\mathrm{ms}^{-1}\right)$\\
    \hline
    0 & 6.77(1) & 0.23(1) & 0 & 8.31(1) & 0.22(2)\\
    $-2 k_L$ & 6.76(1) & 0.19(1) & -$2 k_L$ & 8.30(3) & 0.27(4)\\
    $2 k_L$ & 6.78(1) & 0.22(2) & $2 k_L$ & 8.35(2) & 0.16(6)\\
    \end{tabular}
    \end{ruledtabular}
    \label{Table：S1}
\end{table}

\begin{table}[h!]
    \centering
    \caption{\textbf{Oscillation frequencies and decoherence rates of the Bragg diffracted atomic and molecular populations.} Obtained from the fits in Fig.~$2$d.}
    \begin{ruledtabular}
    \begin{tabular}{c c c c c c}
    \multicolumn{3}{c}{\textbf{Atoms}} & \multicolumn{3}{c}{\textbf{Molecules}}\\
    \hline
    Momentum & Frequency & Decoherence rate & Momentum & Frequency & Decoherence rate\\
    $k$ & $\Omega/2\pi$~(kHz) & $\eta$~$\left(\mathrm{ms}^{-1}\right)$ & $k$ & $\Omega/2\pi$~(kHz) & $\eta$~$\left(\mathrm{ms}^{-1}\right)$\\
    \hline
    0 & 1.60(1) & 0.11(2) & 0 & 2.93(2) & 0.15(1)\\
    $2 k_L$ & 1.59(1) & 0.10(1) & $2 k_L$ & 2.95(1) & 0.14(1)\\
    \end{tabular}
    \end{ruledtabular}
    \label{Table：S2}
\end{table}

\section{III. Phase modulation measurement}
We imprint a phase pattern onto the atomic BEC by pulsing on a moving optical lattice with intensity $I(x,t)=I_0 \sin^2 k_L(x-v_Lt)$, where $v_L=\hbar k_L/m$ is the recoil velocity of the atom. The lattice potential depth is set to $3.63(1)~E_\mathrm{R}$ in all experiments shown in Fig.~$3$. After the Bragg pulse, we shut off the optical lattice and immediately convert the atoms into molecules in $1.5$~ms. Any remaining atoms are blown away with a resonant light pulse. 
When we blow away the atoms, we simultaneously shut off the DMD potential and let the molecules expand in free space for $16$~ms. The dipole trap is left turned on in the $z-$direction so the molecules only expand in the $x$ and $y$ directions. We dissociate the molecules into atoms at the end of this time-of-flight and image the atoms.

The atomic populations for $k=0$ and $2 k_L$ are described by $n_0\propto\cos^2\Omega_\mathrm{a} t/2$ and $n_2\propto\sin^2\Omega_\mathrm{a} t/2$, respectively, while the molecular populations for $k=0$, $2 k_L$ and $4k_L$ are described by $m_0\propto\cos^4\Omega_{0} t/2$,  $m_2\propto\sin^2\Omega_{2} t/2$ and $m_4\propto\sin^4\Omega_{4} t/2$, respectively. A decoherence rate $\eta$ is added to describe the weak decay in the oscillation amplitudes, shown in Fig.~$3$c. The results from the fitting are listed in Table~\ref{Table：S3}.

In Fig.~$4$a, we characterize the phase modulation of the atomic wavefunction $\phi_\mathrm{a}\equiv\varphi_\mathrm{a}(x=\lambda/4)$ at the anti-nodes of the optical lattice. We extract the phase from the diffracted atomic populations according to

\begin{align}
    \cot^2\phi_\mathrm{a}=\frac{n_0}{n_2}.
\end{align}

When determining the phase beyond the first quadrant ($0\leq\phi_\mathrm{a}<\pi/2$) for Fig.~$4$a, we use  $\phi_\mathrm{a}=\Omega_\mathrm{a} t/2$ as a guide to determine the sign of $\cot\phi_\mathrm{a}$ in other quadrants and flip the sign of the cotangent function accordingly.
The correct phase in the second quadrant ($\pi/2\leq\phi_\mathrm{a}<\pi$) is given by $\phi_\mathrm{a} = -\cot^{-1}\sqrt{n_0/n_2} + \pi$.
Due to $\pi$-periodicity of the cotangent, the third quadrant has the same phase as the first and the fourth has the same as the second, and so on. This operation can be summarized in a single formula as $\phi_\mathrm{a}~(\mathrm{mod}~\pi) = (\cot^{-1}\sqrt{n_0/n_2}-\pi/2)\mathrm{sgn}(\sin\Omega_\mathrm{a}t) + \pi/2$, where the function $\mathrm{sgn}(x)=1$ for $x> 0$, $0$ for $x=0$ and $-1$ for $x<0$.
We perform a similar operation to get the molecular phase $\phi_\mathrm{m}~(\mathrm{mod}~\pi)$.

Additionally, there is a $1\%$ offset between the measured atomic and molecular Rabi frequencies due to the systematic drift of the lattice beam intensity in the experiment. To compensate for this systematic effect, we apply a correction factor $\alpha=(\Omega_{0}+\Omega_{4})/2\Omega_\mathrm{a}=0.987$ to the atomic phase $\phi_\mathrm{a}$ based on the measured Rabi frequencies of the molecules and atoms.

Our determination of both the atomic and molecular phase, $\phi_\mathrm{a}$ and $\phi_\mathrm{m}$, is sensitive to the variation of the background, which shows a larger systematic uncertainty of the molecular phase near $\phi_\mathrm{a}=0$ and $\pi/2$.

\begin{table}[h!]
    \centering
    \caption{\textbf{Oscillation frequencies and decay rates of the atomic and molecular populations.} Obtained from the fits in Figs.~$3$c.}
    \begin{ruledtabular}
    \begin{tabular}{c c c c c c}
    \multicolumn{3}{c}{\textbf{Atoms}} & \multicolumn{3}{c}{\textbf{Molecules}}\\
    \hline
    Momentum & Frequency & Decay rate & Momentum & Frequency & Decay rate\\
    $k$ & $\Omega/2\pi$~(kHz) & $\eta$~$\left(\mathrm{ms}^{-1}\right)$ & $k$ & $\Omega/2\pi$~(kHz) & $\eta$~$\left(\mathrm{ms}^{-1}\right)$\\
    \hline
    0 & 2.36(1) & 0.05(1) & 0 & 2.33(1) & 0.08(1)\\
    $2 k_L$ & 2.36(1) & 0.03(1) & $2 k_L$ & 4.66(1) & 0.03(2)\\
     &  &  & $4 k_L$ & 2.33(1) & 0.13(2)\\
    \end{tabular}
    \end{ruledtabular}
    \label{Table：S3}
\end{table}

\section{IV. Optical Lattice Calibration}
Dynamics of a particle with mass $m$ in a one dimensional (1D) lattice potential $V(x)=V_0 \sin^2 k_Lx$ is described by the Schr{\"o}dinger's equation $i \hbar \partial_t\psi=H\psi$, where $k_L=\pi/d$ is the lattice wavenumber, $d=\lambda/2$ is the lattice constant. The Hamiltonian $H$ and the solution of the particle's wavefunction $\psi(x,t)$ are

\begin{align}
H=-\frac{\hbar^2}{2m}\partial_x^2+V_0\sin^2k_Lx. \label{Eq.S2:hamiltonian1DLattice}
\end{align}

\begin{align}
\psi(x,t)=\sum_{j,q}a_{j,q}u_{j,q}(x)e^{-iE_{j,q}t/\hbar}.
\end{align}

\noindent Here $E_{j,q}$ is the eigenenergy of the state (dispersion), $u_{j,q}(x)$ is the eigenfunction (Bloch waves), $j=0,1,2...$ is the Bloch band index, $-k_L/2<q<k_L/2$ is the quasi-momentum, and $a_{j,q}$ is the amplitude. We calculate the eigenstates and the eigenenergies by solving the Mathieu equation

\begin{align}
(\partial_u^2+a-2b\cos 2v)u(v)=0,
\end{align}

\noindent where $a_{j,q}=(E_{j,q}-V_0/2)/E_\mathrm{R}$ and $u_{j,q}(v)$ are the characteristic values and functions, $b=-V_0/4E_\mathrm{R}$, $v=k_Lx$, and $E_\mathrm{R}=\hbar^2k_L^2/2m$ is the recoil energy. The Bloch theorem states that $u_{j,q}=e^{iqx}f_j(x)$, where $f_j(x)=f_j(x+d)$ is a periodic function.

\subsection{Kapitza-Dirac diffraction}
Kapitza-Dirac diffraction occurs when the particle is diffracted to multiple quasi-momentum states. When the particle is initially prepared in the zero momentum state $k=0$, the lattice projects the state to different Bloch bands with quasi-momentum $q=0$. The wavefunction is given by

\begin{align}
\psi(x,t)=\sum_{j=0,2,4...}a_j f_{j}(x)e^{-iE_{j,0}t/\hbar},
\end{align}

\noindent where $a_j$ is the amplitude in the $j$-th excited band. Only even bands are occupied due to the symmetry of the initial state. In the time-of-flight experiment, the particle is scattered to momentum states $k_j=\pm j k_L$. 

The Rabi oscillations in the Kapitza-Dirac diffraction experiment, shown in Figs.~$2$a and $2$b, are dominated by the transition of atoms between the ground and second bands at zero quasi-momentum $q=0$, see Fig.~\ref{fig:S1}a, which is given by 
\begin{align}
\hbar \omega_{\mathrm{K.D.}} = E_{2,q=0}-E_{0,q=0}.
\end{align}
\subsection{Determination of molecular dynamical polarizability}

\begin{figure*}[t!]
    \centering
    \includegraphics[width=\linewidth]{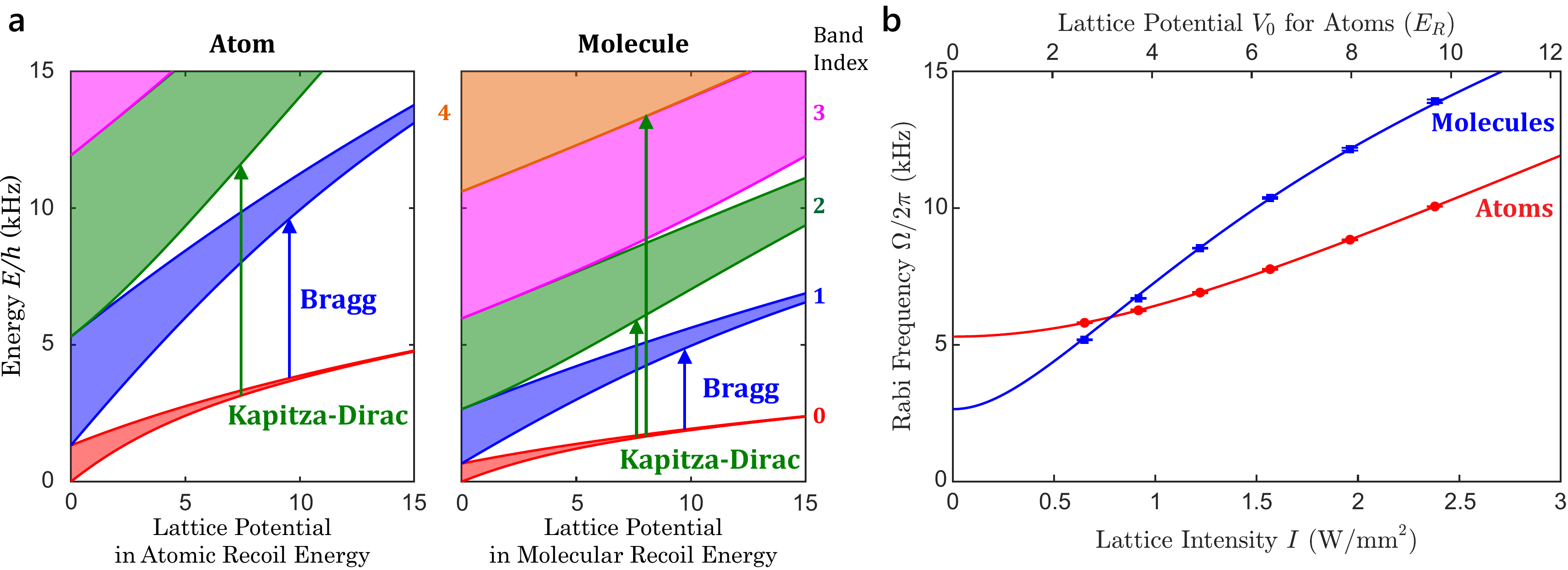}
    \caption{\textbf{Comparison of atomic and molecular polarizabilities.} \textbf{a}, Band structure for atoms and molecules in a 1D lattice in units of atomic and molecular recoil energy, respectively. The energies are obtained from the solution of the Mathieu equation. Band index is indicated by different colors. The lower (upper) edge of the red (blue) band corresponds to the band energy at quasi-momentum $q=0$ while the upper (lower) edge of the red (blue) band corresponds to the band energy at $q=k_L$. The cycle repeats for higher excited bands. In Kapitza-Dirac diffraction, particles are excited from the ground to the \nth{2} excited band at $q=0$. In Bragg diffraction, particles are excited from ground to the \nth{1} excited band at $q=k_L$. \textbf{b}, From the Kapitza-Dirac diffraction measurements, we plot the atomic and molecular Rabi frequencies at various lattice intensities and their corresponding potential depths for atoms (in units of atomic recoil energy $E_R$). The lattice depth is extracted from the Rabi oscillations in which the Rabi frequency corresponds to the energy difference between the ground and \nth{2} excited band. The red solid line is an exact solution to the Mathieu equation for atoms while the blue solid line is a fit based on the solution to the Mathieu equation for molecules with an additional horizontal scaling factor. From fitting the data, we determine the molecular polarizability to be $\alpha_\mathrm{m}=1.95(2)\alpha_\mathrm{a}=k_B\times5.05(5)~\mathrm{nK\cdot cm^2/W}$, where the atomic polarizability is $\alpha_\mathrm{a}=k_B\times2.59~\mathrm{nK\cdot cm^2/W}$.  }
    \label{fig:S1}
\end{figure*}

Molecular dynamical polarizability can be determined from molecular response to the optical lattice potential. We measure the atomic and molecular Rabi oscillation frequencies at various lattice beam intensities from Kapitza-Dirac and Bragg diffraction, see Fig.~\ref{fig:S1}. These frequencies provide precise information on the energy dispersion, from which we extract the light shifts induced by the lattice potential. The lattice potential depth is proportional to the lattice intensity, $V_0=-\alpha_\mathrm{a} I$, where $\alpha_\mathrm{a}$ is the dynamical polarizability of the atoms in a dipole trap and $I$ is the lattice intensity. Since the lattice potential depths for atoms and molecules are different despite the lattice intensity being constant, see examples in Fig.~$2$b and Fig.~\ref{fig:S1}, this indicates the atomic and molecular polarizabilities are not the same. 

We first determine the atomic dynamical polarizability. The light shift on the cesium atoms in the ground state for far-detuned light~\cite{GRIMM200095} can be expressed as the weighted sum of the contributions from the $D_1$ and $D_2$ lines, which gives

\begin{equation}
  V_0=-\left(f_1\frac{3\pi c^2}{\omega_{1}^2}\frac{\Gamma_1}{\omega_{1}^2-\omega_0^2}+f_2\frac{3\pi c^2}{\omega_{2}^2}\frac{\Gamma_2}{\omega_{2}^2-\omega_0^2}\right)I,
    \label{S37}
\end{equation}
where $f_1=1/2.98$ and $f_2=1.98/2.98$ are the line strengths~\cite{RafacTanner1998}, $\omega_{1}=2\pi\times 335.116~\mathrm{THz}$ and $\omega_{2}=2\pi\times 351.726~\mathrm{THz}$ are the resonant frequencies, $\Gamma_1=2\pi\times 4.56~\mathrm{MHz}$ and $\Gamma_2=2\pi\times 5.22~\mathrm{MHz}$ are the natural linewidths of the $D_1$ and $D_2$ lines, respectively, and $c$ is the speed of light. Contributions from other higher excited states are negligible. 

For lattice beams with wavelength $\lambda=1064~\mathrm{nm}$, the lattice frequency is $\omega_0=2\pi\times 281.760~\mathrm{THz}$, and the atomic polarizability is $\alpha_\mathrm{a}=k_\mathrm{B}\times2.59~\mathrm{nK\cdot cm^2/W}$. This value allows us to evaluate the intensity of the optical lattice superimposed on the atoms from the measurement of the lattice potential depth.

We compare the lattice potential depths and their corresponding intensities for both atoms and molecules, see Table \ref{Table:S4}. From the lattice intensity and the induced light shift on the molecules, we determine the molecular polarizability to be
$\alpha_\mathrm{m}=k_\mathrm{B}\times5.05(5)~\mathrm{nK\cdot cm^2/W}$ at $\lambda=1064$~nm. This is $1.95(2)$ times greater than the atomic polarizability $\alpha_\mathrm{a}$. We attribute the slight deviation from the expected ratio of $\alpha_\mathrm{m}/\alpha_\mathrm{a}\approx2$ to the presence of molecular states that can be accessed by $\mathrm{Cs}_2$ molecules.

\begin{table}[t!]
    \centering
    \caption{\textbf{Atomic and molecular Rabi frequencies and their corresponding lattice potential depths} (in units of atomic recoil energy $E_R$) at different lattice intensities shown in Fig.~\ref{fig:S1}.}
    \begin{ruledtabular}
    \begin{tabular}{c c c c c}
    \multicolumn{1}{c}{}& \multicolumn{2}{c}{\textbf{Atoms}} & \multicolumn{2}{c}{\textbf{Molecules}}\\
    \hline
    Lattice intensity & Frequency & Lattice depth & Frequency & Lattice depth\\($\mathrm{W}/\mathrm{mm}^2$) & $\Omega/2\pi$~(kHz) & $V_0~(E_R)$ & $\Omega/2\pi$~(kHz) & $V_0~(E_R)$\\
    \hline
    0.65(1) & 5.81(1) & 2.65(4) & 5.19(2) & 5.06(2)\\
    0.92(2) & 6.27(3) & 3.73(6) & 6.70(1) & 7.09(2)\\
    1.22(1) & 6.92(2) & 4.96(4) & 8.53(2) & 9.64(3)\\
    1.57(1) & 7.77(1) & 6.37(2) & 10.37(4) & 12.43(6)\\
    1.96(1) & 8.85(2) & 7.98(3) & 12.16(5) & 15.53(9)\\
    2.38(1) & 10.06(1) & 9.68(2) & 13.91(6) & 19.03(13)
    \end{tabular}
    \end{ruledtabular}
    \label{Table:S4}
\end{table}

\clearpage
\subsection{Bragg diffraction}
Bragg diffraction occurs when the particle is resonantly coupled to a single quasi-momentum state. The particle dynamics can be well approximated by coherent Rabi flopping between the initial and the excited states. 

In our experiment, Bragg diffraction is realized by imposing an optical lattice $V(x,t)=V_0 \sin^2 k_L(x-v_Lt)$ moving at the recoil velocity $v_L=\hbar k_L/m$ of the particle. The lattice resonantly excites stationary particles to momentum $k=2k_L$ with negligible populations in other states. In the moving frame, the dynamics is described by the same stationary Hamiltonian Eq.~(\ref{Eq.S2:hamiltonian1DLattice}) and the particles initially moving with momentum $-k_L$ in the ground band $u_{0,q=-k_L}$ are driven to $u_{1,q=k_L}$ in the first excited band with momentum $k_L$.

The Rabi oscillations in Figs.~$2$c and $2$d are dominated by the transition between the two momentum states $\ket{k=0}$ and $\ket{k=2k_L}$ in the lab frame. Using $E_{0,-k_L}=E_{0,k_L}$, we obtain the Rabi frequency $\omega_{\mathrm{B}}$ of Bragg scattering as

\begin{align}
\hbar \omega_{\mathrm{B}} = E_{1,k_L}-E_{0,k_L}.
\end{align}

Bragg scattering induces coherent oscillations between the two states.  The wavefunction in the lab frame can be written in the momentum basis, polar form and spin basis with the basis vectors $\ket{0}\equiv\ket{k=0}$ and $\ket{1}\equiv\ket{k=2k_L}$ as:

\begin{align}
\psi(x,t)&=\cos\tau - i\sin\tau e^{2ik_L x}\label{eqS91:atom_wf}\\
        &\equiv\sqrt{\rho(x,t)}e^{i\varphi(x,t)}\label{eqS92:atom_wf}\\
        &\equiv e^{-i\hat{\sigma}_x \tau}|0\rangle =\cos \tau |0\rangle -i \sin \tau |1\rangle, \label{eqS9:atom_wf}
\end{align}

\noindent where $e^{-i\hat{\sigma}_x \tau}$ is the spin rotation operator that describes the dynamics between the two states, $\hat{\sigma}_x$ is the Pauli$-x$ matrix, $\rho(x,t)$ is the particle density, $\varphi(x,t)$ is the phase modulation and $\tau=\Omega t/2$. The population in the zero-momentum state is $n_0 = |\langle 0|\psi\rangle|^2=\cos^2 \tau$ and  $n_2=|\langle2k_L|\psi\rangle|^2=\sin^2 \tau$ in the excited state. We evaluate the phase modulation amplitude $\phi\equiv\varphi (x=\lambda/4)$ at the anti-nodes of the lattice. The phase modulation amplitude is also equivalent to the spin rotation angle $\phi=\tau$.
\newpage
\section{V. Reaction phase $\varphi^{}_\mathrm{R}=2\varphi_\mathrm{a}-\varphi_\mathrm{m}$}
Reactive coupling between atomic and molecular Bose-Einstein condensates can be modeled in the dilute gas limit by two-body Feshbach coupling

\begin{align}
\hat{H_2}=\hbar \gamma\hat{\psi}_\mathrm{m}^{\dagger}\hat{\psi}_\mathrm{a}^2+\mathrm{H.c.},
\end{align}

\noindent where $\gamma$ is the strength of Feshbach coupling, $\hat{\psi}_\mathrm{m}$ and $\hat{\psi}_\mathrm{a}$ are the atomic and molecular field operators, respectively. 
This Hamiltonian conserves the total particle number $N_0=N+2M$. 
In systems with high densities, three-body coupling $\hat{H_3}=\hbar\gamma_3\hat{\psi}_\mathrm{m}^{\dagger}\hat{\psi}_\mathrm{a}^{\dagger}\hat{\psi}_\mathrm{a}^3+\mathrm{H.c.}$ dominates \cite{Zhang2023}. 
In the following, we consider $\hat{H_2}$ for simplicity. Generalization of the following calculation to $\hat{H_3}$ is straightforward.

The equations of motion for the atomic wavefunction $\hat{\psi}_\mathrm{a}$ and the molecular wavefunction $\hat{\psi}_\mathrm{m}$, and their respective populations $\hat{N}=\hat{\psi}_\mathrm{a}^{\dagger}\hat{\psi}_\mathrm{a}$ and $\hat{M}=\hat{\psi}_\mathrm{m}^{\dagger}\hat{\psi}_\mathrm{m}$ can be obtained from the Heisenberg equation as

\begin{align}
\partial_t\hat{\psi}_\mathrm{m}&=-i\gamma\hat{\psi}_\mathrm{a}^2 \\
\partial_t\hat{\psi}_\mathrm{a}&=-2i\gamma\hat{\psi}_\mathrm{a}^{\dagger}\hat{\psi}_\mathrm{m} \\
\partial_t\hat{M}&=-\frac12\partial_t\hat{N}=\frac{i}{\hbar}[\hat{H_2},\hat{M}]=i\gamma(\hat{\psi}_\mathrm{a}^{\dagger}{}^2\hat{\psi}_\mathrm{m}-\hat{\psi}_\mathrm{m}^{\dagger}\hat{\psi}_\mathrm{a}^2).
\end{align}
 
For large atomic populations $N\equiv\langle\hat{N}\rangle\gg1$ and molecular populations $M\equiv\langle\hat{M}\rangle\gg 1$, we assume both fields are described by coherent states  $\ket{\psi}=\ket{c_\mathrm{a}}\otimes\ket{c_\mathrm{m}}$, where $c_\mathrm{a}=\sqrt{N}e^{i\varphi_\mathrm{a}}$ and $c_\mathrm{m}=\sqrt{M}e^{i\varphi_\mathrm{m}}$ are complex eigenvalues. The equations of motion for populations $N$ and $M$ and the reaction phase $\varphi^{}_\mathrm{R}=2\varphi_\mathrm{a}-\varphi_\mathrm{m}$ are given by

\begin{align*}
\dot{M}&=-\frac{\dot{N}}{2}
=2\gamma N\sqrt{M}\sin\varphi^{}_\mathrm{R}\\
\dot{\varphi}_\mathrm{R}&=\gamma\frac{N-4M}{\sqrt{M}}\cos\varphi^{}_\mathrm{R}.
\end{align*}

Here we consider a few limiting cases. Atom-molecule equilibrium is reached when atom and molecule numbers are stationary with population $N=2N_0/3$ and $M=N_0/6$. In this case, the reaction phase is zero $\varphi^{}_\mathrm{R}=0$, which suggests phase doubling $\varphi_\mathrm{m}=2\varphi_\mathrm{a}$.

Near equilibrium, we can linearize the equation to leading order in $\delta M\equiv M-N_0/6$ and $\delta N\equiv N-2N_0/3=-2\delta M$. The populations and phase oscillate according to

\begin{align*}
\delta M(t)&=\delta M(0)\cos\Omega t\\
&=-\frac12\delta N(t)=\frac{2N_0^{3/2}}{\sqrt{3}}\varphi^{}_\mathrm{R}(t),
\end{align*}

\noindent where the oscillation frequency $\Omega=\sqrt{8N_0}\gamma$ increases with sample size due to Bose enhancement~\cite{Zhang2023}. This result shows that a positive reaction angle $\varphi^{}_\mathrm{R}$ is linked to excess molecules. In addition, the reaction angle near the equilibrium $\varphi^{}_\mathrm{R}(t)\propto N_0^{-3/2}\delta M(t)$ approaches zero in the thermodynamic limit $N_0\gg 1$. Thus $\varphi_\mathrm{m}\approx2\varphi_\mathrm{a}$ remains a very good approximation near equilibrium. 

When the system is far from equilibrium, for example, at the beginning of our experiment when the sample is dominated by atoms with $N\gg 2M\gg1$, atoms are paired into molecules. The reaction phase $\varphi^{}_\mathrm{R}\approx\pi/2$ is the only stationary solution and the molecular number $M(t)=(\gamma N_0t)^2$ increases quadratically with time. Here we have $2\varphi_\mathrm{a}=\varphi_\mathrm{m}+\pi/2$.

In the opposite limit when the sample is dominated by molecules $M\gg N/2\gg1$, Feshbach coupling dissociates molecules back to atoms. Here the reaction phase $\varphi^{}_\mathrm{R}\approx-\pi/2$ is the only stationary solution and the atom number $N(t)=N(t)e^{\Omega t}$ increases exponentially due to stimulated reactive dissociation. Thus we have $2\varphi_\mathrm{a}=\varphi_\mathrm{m}-\pi/2$. We summarize these results in Table~\ref{Table:S5}. 

\begin{table}
    \centering
    \caption{\textbf{Summary of the coherent state dynamics of an atom-molecule BEC with resonant Feshbach coupling.}}
    \begin{ruledtabular}
    \begin{tabular}{lccc}
         & Atom number & Molecule number  & Reaction phase \\
        Condition & $N$ & $M$  & $\varphi^{}_\mathrm{R}=2\varphi_\mathrm{a}-\varphi_\mathrm{m}$\\
        \hline
         Equilibrium $N=4M$ &  $\frac{2}{3}N_0$ & $\frac16 N_0$   & 0 \\
         Near equilibrium $N\approx 4M$& $\frac{2}{3}N_0+\Delta\cos\Omega t$ & $\frac16 N_0-\frac{\Delta}2\cos\Omega t$  & $\approx 0$ \\
         Atom dominated $N\gg 2M$& $\approx N_0$ & $(\gamma N_0 t)^2$ & $\pi/2$ \\
         Molecule dominated $M\gg N/2$& $e^{\Omega t}$ & $\approx \frac12 N_0$ & $-\pi/2$\\
    \end{tabular}
    \end{ruledtabular}
    \label{Table:S5}
\end{table}

\section{VI. Molecule Synthesis in Two-component BEC\lowercase{s}: Collective Spin Picture}
When particles are diffracted by a Bragg pulse, it is instructive to represent atoms and molecules in different momentum states using angular momentum representation. For our experiment in Fig.~$3$, the full many-body Hamiltonian can be written as

\begin{align}
    H&=\sum_\kappa\epsilon_\kappa(\hat{a}_\kappa^{\dagger }\hat{a}_\kappa+\frac12\hat{m}^{\dagger }_\kappa \hat{m}_\kappa )+\hbar\frac{\Omega_\mathrm{a}}2\sum_\kappa (\hat{a}_{\kappa +2}^{\dagger }\hat{a}_\kappa +\hat{a}_\kappa ^{\dagger }\hat{a}_{\kappa +2})+\gamma\sum_{\kappa _1,\kappa _2}(\hat{m}^{\dagger }_{\kappa _1+\kappa _2}\hat{a}_{\kappa _1}\hat{a}_{\kappa _2}+\mathrm{H.c.}).
\end{align}

\noindent Here $\hat{a}_\kappa ^\dagger$ and $\hat{a}_\kappa$ are the atomic field operators, $
\hat{m}_\kappa ^\dagger$ and $\hat{m}_\kappa$ are the molecular field operators, $\kappa _1$ and $\kappa _2$ are dimensionless momenta in units of the lattice momentum $k_L$, $\epsilon_\kappa =\kappa ^2 E_\mathrm{R}$ is the kinetic energy of an atom, $\Omega_\mathrm{a}$ is the atomic Rabi frequency induced by the Bragg coupling, and $\gamma$ is the atom-molecule coupling strength.

The Bragg pulse resonantly couples atoms in two momentum modes $a_0$ and $a_2$, and pairing these atoms produces molecules in three momentum modes $m_0$, $m_2$ and $m_4$ due to momentum conservation. In the Hilbert space of these modes, we use Schwinger's angular momentum representation to describe the collective excitations of atoms: $J_x=\frac12(\hat{a}_0^{\dagger }\hat{a}_2+\hat{a}_2^{\dagger }\hat{a}_0)$, $J_y=\frac{-i}2(\hat{a}_0^{\dagger }\hat{a}_2-\hat{a}_2^{\dagger }\hat{a}_0)$, $J_z=\frac12(\hat{a}_0^{\dagger }\hat{a}_0-\hat{a}_2^{\dagger }\hat{a}_2)$. 
In the interaction picture, the Hamiltonian reduces to

\begin{align}
    H&\equiv -\hbar\Delta \hat{m}_2^{\dagger }\hat{m}_2+\hbar\Omega_\mathrm{a}J_x+\hbar\gamma(\hat{m}^{\dagger }_0\hat{a}_0^2+2\hat{m}^{\dagger}_2\hat{a}_0\hat{a}_2+\hat{m}^{\dagger }_4\hat{a}_2^2)+\mathrm{H.c}..
\end{align}

\noindent In our experiment, the Bragg pulse resonantly couples two atoms with identical momentum $k$ to a molecule with momentum $2k$. However, coupling two atoms with momenta $0$ and $2k_L$ into a molecule, described by $\hat{m}^{\dagger }_2\hat{a}_0\hat{a}_2$, is an off-resonant process with detuning $\hbar\Delta=2E_R$ since the molecule with $2k_L$ has a lower kinetic energy than the atoms by two atomic recoil energies $2E_R$. This detuning is included in the first term of the Hamiltonian.

Atoms are initially prepared in the BEC with zero momentum $k=0$, collectively described by $|i\rangle\equiv|J=N/2, m_J=-J\rangle$. The Bragg pulse rotates the state by $R=e^{-iJ_x\Omega_\mathrm{a} t}$. The pulse is followed by the Feshbach coupling $\gamma$, which pairs the atoms into molecules in three different momentum modes. From the Heisenberg picture, the equation of motion and the production of molecules in different states at very short times are given by

\begin{align}
   i \partial_t \hat{m}_0&=-[H,\hat{m}_0]=\gamma \hat{a}_0^2 &&\Rightarrow M_0\approx\langle \hat{a}_0^\dagger \hat{a}_0^\dagger \hat{a}_0 \hat{a}_0\rangle  \gamma^2t^2 = N_0(N_0-1)\gamma^2t^2 \\
    i \partial_t \hat{m}_2&=-[H,\hat{m}_2]=-\Delta \hat{m}_2+2 \gamma \hat{a}_0\hat{a}_2 &&\Rightarrow M_2\approx4\langle \hat{a}_0^\dagger \hat{a}_2^\dagger \hat{a}_2 \hat{a}_0 \rangle\gamma'^2 t^2=4 N_0N_2\gamma_2^2t^2\\
    i \partial_t \hat{m}_4&=-[H,\hat{m}_4]=\gamma \hat{a}_2^2 &&\Rightarrow M_4\approx \langle \hat{a}_2^\dagger \hat{a}_2^\dagger \hat{a}_2 \hat{a}_2\rangle\gamma^2 t^2= N_2(N_2-1)\gamma^2t^2,
\end{align}

\noindent where $N_0=\langle \hat{a}_0^\dagger \hat{a}_0\rangle$ and $N_2=\langle \hat{a}_2^\dagger \hat{a}_2\rangle$ are the atomic populations in $k=0$ and $2k_L$ modes, and $M_0\equiv\langle \hat{m}_0^{\dagger }\hat{m}_0\rangle$, $M_2\equiv\langle \hat{m}_2^{\dagger }\hat{m}_2\rangle$ and  $M_4\equiv\langle \hat{m}_4^{\dagger }\hat{m}_4\rangle$ are the molecular populations in $k=0$, $2k_L$, and $4k_L$ modes, and $\gamma_2=\sqrt{g^{(2)}}\gamma'$ is the off-resonance atom-molecule coupling. We introduce the second order correlation function to factorize the correlation of the atomic fields as 

\begin{align}
   g^{(2)}=\frac{\langle \hat{a}_0^\dagger \hat{a}_2^\dagger \hat{a}_2 \hat{a}_0\rangle}{\langle \hat{a}_0^\dagger \hat{a}_0\rangle\langle \hat{a}_2^\dagger \hat{a}_2\rangle}.
\end{align}

For Bragg-diffracted atomic BEC, we expect the correlation function to be $g^{(2)}=(N-1)/N\approx 1$, where $N\gg1$ is the atom number. 

\section{VII. Molecule synthesis of diffracted atoms: Mean-field picture}
To compare our measurement results with theory, we model the atomic and molecular wavefunctions based on mean-field wavefunctions. Given the wavefunction of a Bragg-diffracted atom is $\psi_\mathrm{a}(x)\equiv \psi_\mathrm{a_0}+\psi_\mathrm{a_2}e^{2ik_Lx}=\cos\tau-i\sin\tau e^{2ik_Lx}$, where $\tau=\Omega_\mathrm{a}t/2$ and the molecular wavefunction is $\psi_\mathrm{m}(x)\equiv\psi_\mathrm{m_0}+\psi_\mathrm{m_2}e^{2ik_Lx}+\psi_\mathrm{m_4}e^{4ik_Lx}$. The results of Eqs.~(20)–(22) can be translated into the probabilities of finding the molecules in three momentum modes, which give us

\begin{align}
m_0&=|\psi_\mathrm{m_0}|^2\propto \gamma^2|\psi_\mathrm{a_0}|^4=\gamma^2\cos^4\tau\\
m_2&=|\psi_\mathrm{m_2}|^2\propto 4\gamma_2^2|\psi_\mathrm{a_0}|^2|\psi_\mathrm{a_2}|^2=\gamma_2^2\sin^2 2\tau\\
m_4&=|\psi_\mathrm{m_4}|^2\propto \gamma^2|\psi_\mathrm{a_4}|^4=\gamma^2\sin^4\tau.
\end{align}

\noindent 
These are the fit functions we used in Fig.~$3$b to describe the molecular data. Based on the equations above, we model the molecular wavefunction as

\begin{align}
    \psi_\mathrm{m}(x,t)\equiv\langle x|\psi_\mathrm{m}\rangle &=A_\mathrm{m} (\gamma\cos^2\tau-2i\gamma_2\sin\tau\cos\tau e^{2ik_Lx}-\gamma\sin^2\tau e^{4ik_Lx}) \label{eqS25:psi_mol}\\
    &\equiv\sqrt{\rho_m(x,t)}e^{i\varphi_\mathrm{m}(x,t)},
\end{align}

\noindent where $A_\mathrm{m}=[\gamma^2+(\gamma_2^2-\gamma^2/2)\sin^2 2\tau]^{-1/2}$ is the normalization constant. Here we have introduced the density $\rho_m(x,t)$ and phase $\varphi_\mathrm{m}(x,t)$ of the molecular wavefunction. The probabilities of molecules with momenta $0$, $2k_L$ and $4k_L$ are 

\begin{align}
    m_0&=A_\mathrm{m}^2\gamma^2 \cos^4\tau \label{eqS27:m0}\\
    m_2&=A_\mathrm{m}^2\gamma_2^2 \sin^22\tau\\
    m_4&=A_\mathrm{m}^2\gamma^2 \sin^4\tau. \label{eqS29:m4}
\end{align}

\noindent In particular, the population of $m_2$ reaches its maximum of $m_2(\tau=\pi/4)=\gamma_2^2/(\gamma_2^2+\gamma^2/2)$ after a $\pi/2-$pulse on the atoms. From the measurement of $m_2=0.58(1)$, we determine $\gamma_2/\gamma=0.82(3)$. The smaller value of sum-frequency coupling constant $\gamma_2$ compared to Feshbach coupling $\gamma$ can indicate anti-correlations between the populations in $k=0$ and $2k_L$ modes. In addition, the detuning of sum-frequency generation $\delta E=-2E_R=-h\times 2.7$~kHz is non-negligible compared to the coupling strength $\Gamma=\delta\mu\Delta B=h\times 6.3(5)$~kHz of the $g-$wave Feshbach resonance, calculated from the resonance width $\Delta B=8.3 (5)$~mG and relative magnetic moment $\delta \mu = h\times0.76(3)~\mathrm{MHz}/\mathrm{G}$ \cite{Zhang2023}. The atom-molecule conversion efficiency for sum-frequency generation is thus suppressed. Using the Lorentzian lineshape as an estimation, we obtain a lower conversion efficiency of $1/(1+4\delta E^2/\Gamma^2)\approx0.59(4)$, leading to $\gamma_2/\gamma=\sqrt{0.59(4)}=0.77(3)$, which is consistent with our observed value $\gamma_2/\gamma=0.82(3)$.

The phase modulation of the molecular wavefunction evaluated at the antinodes of the optical lattice $\phi_\mathrm{m}\equiv \varphi_\mathrm{m}(x=\lambda/4)$ gives

\begin{align}
    \phi_\mathrm{m}=\cot^{-1}\frac{\sqrt{m_0}-\sqrt{m_4}}{\sqrt{m_2}}=\cot^{-1}\frac{\gamma}{\gamma_2}\cot2\tau.\label{eqS34:mol_phase_tau}
\end{align}

We may express the molecular phase $\phi_\mathrm{m}$ in terms of the atomic phase $\phi_\mathrm{a}=\tau$ using Eq.~(\ref{eqS34:mol_phase_tau}), which gives

\begin{align}
    \phi_\mathrm{m}&=\cot^{-1}\frac{\gamma}{\gamma_2}\cot2\phi_\mathrm{a}\\
          &=2\phi_\mathrm{a}-\epsilon \sin 4\phi_\mathrm{a}+\mathcal{O}(\epsilon ^2),
\end{align}

\noindent where we have expanded the nonlinear parameter $\epsilon =(1-\gamma_2/\gamma)/2$ to leading order. 
This expression shows that phase doubling $\phi_\mathrm{m}=2\phi_\mathrm{a}$ exactly occurs when the atom-molecule coupling is momentum insensitive $\gamma_2=\gamma$ and the correlator is $g^{(2)}=1$.

\section{VIII. Parity, coherence and entanglement of the molecular state}

Consider $N$ spins with angular momenta $\vec{s_j}=\frac{\hbar}{2}\vec{\sigma_j}$ for the $j-$th spin, where $\vec{\sigma}$ is the Pauli vector. 
We define the spin coherence operators as $\hat{C}_{kk}=\prod_j^N\sigma_{j,k}$, where $k=x,y$ and $z$.

For $N=2$ spins in a product state $|\psi\rangle=|1\rangle\otimes |2\rangle$, we can define the Bloch vectors $\vec{b}_j=\langle j|\vec{\sigma}|j\rangle=(b_{j,x}, b_{j,y}, b_{j,z})$, where $j=$1, 2 refers to the spins and $|\vec{b}_j|=1$. The combinations of the correlators for any product state are bound as

\begin{align}
    \pm C_{xx}\pm C_{yy} \pm C_{zz} & \leq |C_{xx}|+|C_{yy}|+|C_{zz}| \\
    |C_{xx}|+|C_{yy}|+|C_{zz}|&=|\langle\psi|\sigma_{1,x}\sigma_{2,x}|\psi\rangle|+|\langle\psi|\sigma_{1,y}\sigma_{2,y}|\psi\rangle|+|\langle\psi|\sigma_{1,z}\sigma_{2,z}|\psi\rangle|\\
    &=|b_{1,x}b_{2,x}|+|b_{1,y}b_{2,y}|+|b_{1,z}b_{2,z}|\\
    &\leq \sqrt{b^2_{1,x}+b^2_{1,y}+b^2_{1,z}}  \sqrt{b^2_{2,x}+b^2_{2,y}+b^2_{2,z}}=1.
\end{align} 

In our experiment, two atoms are paired into a molecule. If the state during the reaction is a product state of the atoms with identical Bloch vectors $\vec{b}_1=\vec{b}_2=(b_{x}, b_{y}, b_{z})$, the parity $C_{zz}$ is constrained as

\begin{align}
C_{zz}&=\langle\psi|\sigma_{1,z}\sigma_{2,z}|\psi\rangle=b_z^2\\
    0&\leq b_z^2\leq b_{x}^2+b_y^2+b_z^2=1 \\
    \Rightarrow 0&\leq C_{zz}\leq 1.\label{eqS48:Czz_inequality}
\end{align} 

Thus our parity measurement covering a range $-0.15\lesssim C_{zz}<1$ indicates that the reacting atoms cannot be described by a product state of two atoms. The strongest violation occurs when the atoms are prepared in the equal superposition state of $|0\rangle$ and $|2k_L\rangle$ after the $\pi/2$ pulse.

For an entangled state, the above inequality Eq.~\eqref{eqS48:Czz_inequality} can be violated. 
For example, for Bell states $|\Phi^{\pm}\rangle=(|0,0\rangle\pm|1,1\rangle)/\sqrt{2}$ and $|\Psi^{\pm}\rangle=(|0,1\rangle\pm|1,0\rangle)/\sqrt{2}$, the combinations $\hat{W}_{\Phi^\pm}=\pm \hat{C}_{xx}\mp \hat{C}_{yy}+\hat{C}_{zz}-\hat{1}$ and $\hat{W}_{\Psi^\pm}=\pm \hat{C}_{xx}\pm \hat{C}_{yy}-\hat{C}_{zz}-\hat{1}$ serve as entanglement witnesses, which are greater than zero for the corresponding Bell states

\begin{align}
    \langle\Phi^\pm|\hat{W}_{\Phi^\pm}|\Phi^\pm\rangle=\langle\Psi^\pm|\hat{W}_{\Psi^\pm}|\Psi^\pm\rangle=2,
\end{align}
 and any other non-separable states. On the other hand, the witnesses give nonpositive values $W_{\Phi^\pm},W_{\Psi^\pm}\leq 0$ for product states. In our experiment, $\ket{k=0}$ corresponds to $\ket{0}$ and $\ket{k=2k_L}$ and corresponds to $\ket{1}$ in the Bell state notation. Possible witness values are summarized in Table~\ref{tab:witnesses}.

\begin{figure*}[t!]
    \centering \includegraphics[width=0.8\linewidth]{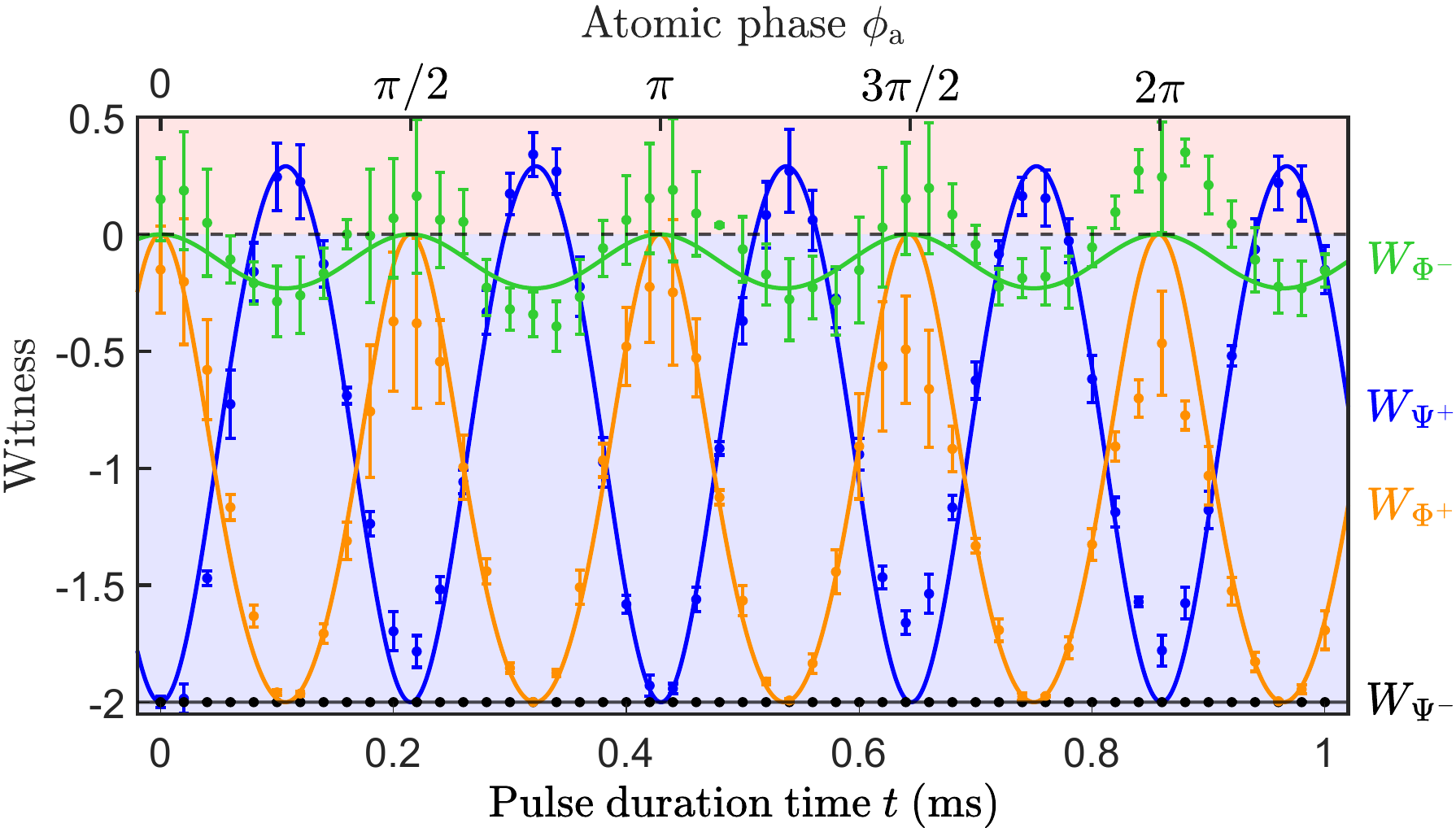}
    \caption{\textbf{Entanglement witnesses.} We plot $W_{\Phi^+}$ (orange), $W_{\Phi^-}$ (green), $W_{\Psi^+}$ (blue), and $W_{\Psi^-}$ (black) as a function of pulse duration time $t$. The red and blue areas represent non-separable and separable states, respectively. $W_{\Psi^+}$ exceeds the upper limit of zero at $\phi_a=\pi/4$, $3\pi/4$, $5\pi/4$, $7\pi/4$, ..., reaching a maximum of $W_{\Psi^+}=0.29(2)$ from the fit, thus indicating entanglement. $W_{\Phi^-}$ appears to exceed zero at $\phi_\mathrm{a}=0, \pi/2$, $\pi$, $3\pi/2$, ..., due to large relative uncertainties of the populations close to zero in Eq.~(\ref{eqS56:Witness_psi-}). Furthermore, the majority of atoms occupy a single momentum state (either $\ket{0}$ or $\ket{2k_L}$) at $\phi_\mathrm{a}=0, \pi/2$, $\pi$, $3\pi/2$, ..., and thus are unable to generate an entangled state.
    }
\end{figure*}

\begin{table}[h!]
    \centering
    \caption{\textbf{Bell state witnesses for different two-spin states.} The last column shows the expected witnesses at $\tau=\pi/4$ for the ideal case when Feshbach couplings for sum frequency and second harmonic generation are equal.}
    \hspace{0.1cm}
    \begin{ruledtabular}
    \begin{tabular}{l c c c c c c c}
         & $|\Phi^{+}\rangle$ & $|\Phi^-\rangle$ & $|\Psi^{+}\rangle$ & $|\Psi^-\rangle$ & product state & \textbf{our exp.} & $\gamma_2=\gamma$\\
         & $\frac{|00\rangle+|11\rangle}{\sqrt{2}}$ & $\frac{|00\rangle-|11\rangle}{\sqrt{2}}$ & $\frac{|01\rangle+|10\rangle}{\sqrt{2}}$ & $\frac{|01\rangle-|10\rangle}{\sqrt{2}}$ & &($\mathbf{\tau=\pi/4}$) &\\
         \hline
        $W_{\Phi^{+}}$ & 2 & -2 & -2 & -2 & $\leq 0$ & \textbf{-2.0(1)} & -2 \\
        $W_{\Phi^-}$ & -2 & 2 & -2 & -2 & $\leq 0$ & \textbf{-0.23(2)} & -2/3\\
        $W_{\Psi^{+}}$ & -2 & -2 & 2 & -2 & $\leq 0$ & \textbf{0.29(2)} & 2/3\\
        $W_{\Psi^-}$ & -2 & -2 & -2 & 2 & $\leq 0$ & \textbf{-2.00(2)} & -2
    \end{tabular}
    \end{ruledtabular}
    \label{tab:witnesses}
\end{table}

We may evaluate the witnesses for the momentum states of reacting atoms from the molecular populations, see Eqs.~(\ref{eqS25:psi_mol}-\ref{eqS29:m4}), which yields

\begin{align}
    C_{xx}&=m_2-2\sqrt{m_0m_4}\\
    C_{yy}&=m_2+2\sqrt{m_0m_4} \\
    C_{zz}&=m_0-m_2+m_4\\
    W_{\Psi^{+}}&=3m_2-m_0-m_4-1=-2C_{zz}\label{eqS50:Wphi+}\\ 
    W_{\Psi^-}&=-m_0-m_2-m_4-1=-2\\
    W_{\Phi^{+}}&=m_0-m_2+m_4-4\sqrt{m_0m_4}-1\\
    W_{\Phi^-}&=m_0-m_2+m_4+4\sqrt{m_0m_4}-1.\label{eqS56:Witness_psi-}
\end{align}

In our experiment, after $\pi/2$ pulse on the atoms we observe $m_2=0.58(1)$ and $m_0=m_4=0.21(1)$ on the molecules. We evaluate the entanglement witness $W_{\Psi^{+}}$ using Eq.~\eqref{eqS50:Wphi+}. From fitting the witnesses Eqs.~(\ref{eqS50:Wphi+}-\ref{eqS56:Witness_psi-}) with Eqs.~(\ref{eqS27:m0}-\ref{eqS29:m4}), using amplitude and frequency as free parameters, we obtain $W_{\Psi^{+}}=0.29(2)$ at $\tau=\pi/4,\ 3\pi/4,\ldots$, which suggests the atomic pair wavefunction during the reaction is not separable and is more aligned with $|\Psi^{+}\rangle$ compared to other Bell states with lower values of $W_{\Psi^-}=-2$, $W_{\Phi^{+}}=-2$ and $W_{\Phi^-}=-0.23(2)$.

For a $\pi/2$-pulse, the witness $W_{\Psi^{+}}$ from Eq.~\eqref{eqS50:Wphi+} can be expressed in terms of ratio $\gamma_2/\gamma$ as
\begin{align}
    W_{\Psi^{+}} (\tau = \pi/4) = \frac{2\gamma_2^2-\gamma^2}{\gamma_2^2+\gamma^2/2}.
\end{align}

By using Eq.~\ref{eqS50:Wphi+}  and $m_0(\tau=\pi/4)=m_4(\tau=\pi/4)$, we can express the wavefunction of the reacting pair after $\pi/2$-pulses as
\begin{align}
    \frac{1}{2}\sqrt{2-W_{\Psi^+}}\ket{\Phi^-} - \frac{i}{2}\sqrt{2+W_{\Psi^+}}\ket{\Psi^+}.
\end{align}

\bibliographystyle{naturemag}
\bibliography{filteredbibliography}